# Viscoroute 2.0: a tool for the simulation of moving load effects on asphalt pavement

**Chabot* A. — Chupin* O. — Deloffre* L. — Duhamel** D.**

**Laboratoire Central des Ponts et Chaussées*
*Division Infrastructures et Matériaux pour Infrastructures de Transport*
*BP 4129 – 44341 Bouguenais Cedex - France*

*armelle.chabot@lcpc.fr*
*olivier.chupin@lcpc.fr*
*lydie.deloffre@lcpc.fr*

***Ecole Nationale des Ponts et Chaussées – Université Paris-Est*
*UR Navier*
*6 et 8 avenue Blaise Pascal – cité Descartes – Champs sur Marne*
*77455 Marne la Vallée – France*

*duhamel@enpc.fr*

*ABSTRACT. As shown by strains measured on full scale experimental aircraft structures, traffic of slow-moving multiple loads leads to asymmetric transverse strains that can be higher than longitudinal strains at the bottom of asphalt pavement layers. To analyze this effect, a model and a software called ViscoRoute have been developed. In these tools, the structure is represented by a multilayered half-space, the thermo-viscoelastic behaviour of asphalt layers is accounted by the Huet-Sayegh rheological law and loads are assumed to move at constant speed. First, the paper presents a comparison of results obtained with ViscoRoute to results stemming from the specialized literature. For thick asphalt pavement and several configurations of moving loads, other ViscoRoute simulations confirm that it is necessary to incorporate viscoelastic effects in the modelling to well predict the pavement behaviour and to anticipate possible damages in the structure.*

## 1. Introduction

The French design method (SETRA-LCPC, 1997) is based on the axisymmetric Burmister multilayer model (1943) which is used in the ALIZE software (www.lcpc.fr). In this 2D static model, each layer has a homogeneous and an elastic behaviour. Viscoelastic effects due to asphalt materials are taken into account only through an equivalent elastic modulus which is determined from complex modulus tests. The equivalent elastic modulus is set for a temperature of 15°C (average temperature in France) and a frequency of 10 Hz. The latter is supposed to be equivalent to a vehicle speed of 72 km/h (average speed of vehicles in France). Semi-analytical calculations relying on the Burmister formalism yield a relatively good approximation of stress and strain fields for pavements under heavy traffic. This is especially true for base courses composed of classical materials. On the contrary, moving load effects and the thermo-viscoelastic behaviour of asphalt materials must be accounted to well represent the behaviour of flexible pavements under low traffic and at high temperatures. The analysis of damages triggered by slow and heavy multiple loads also requires these elements to be considered.

Since early works by Sneddon (1952), the 3D theoretical response of a half-space under a moving load with static and dynamic components has been largely investigated. In the pavement framework, three-dimensional Finite Element-based models have been proposed (e.g. Heck *et al.*, 1998; Elseifi *et al.*, 2006). However, these models may be hard to manipulate, and to offer fast alternative tools, semi-analytical methods are still developed (Hopman, 1996; Siddharthan *et al.*, 1998). In France, at LCPC, (Duhamel *et al.*, 2005) developed such a 3D model which is implemented in the ViscoRoute software. This program directly integrates the viscoelastic behaviour of asphalt materials through the Huet-Sayegh model which is particularly well-suited for the modelling of asphalt overlays (Huet, 1963; 1999; Sayegh, 1965). The ViscoRoute kernel has been validated by comparison to analytical solutions (semi-infinite medium), Finite Element simulations (multilayered structure) and experimental results coming from the LCPC Pavement Fatigue Carrousel (Duhamel *et al.*, 2005). Previous simulations performed with ViscoRoute have confirmed that the equivalent elastic modulus and the time-frequency equivalence assumed by the French design method can be used only for base courses of medium thickness. For bituminous wearing courses and thick flexible pavement structures, especially for aircraft structures solicited by several heavy wheels and at low traffic, it is necessary to develop other concepts (Chabot *et al.*, 2006).

First, the aim of this article is to present the ViscoRoute 2.0 software. To the contrary of the first version (Duhamel *et al.*, 2005), ViscoRoute 2.0 enables to take into account, directly into the computation kernel, multi-loading cases. The possibility of using elliptical-shaped loads has also been added and the Graphic User Interface (GUI) has been completely re-written in the Python language.

Then, this paper highlights multi-loading effects on viscoelastic pavements. The results presented herein for twin wheels, tandem and tridem load cases might also be useful in the context of the new generation of European trucks which is currently under study.

## 2. Problem description

The pavement structure is assimilated to a semi-infinite multilayered medium. It is composed of *n* horizontal layers that are piled up in the z-direction. $e^i$ and $\rho^i$ denote the thickness and the density of layer $i$ $\left(i \in \{1,n\}\right)$, respectively. The structure is solicited by one or several moving loads that move in the x-direction with a constant speed, *V*. The load pressure can be applied in any of the three directions, at the free surface (z=0) of the medium (Figure 1).

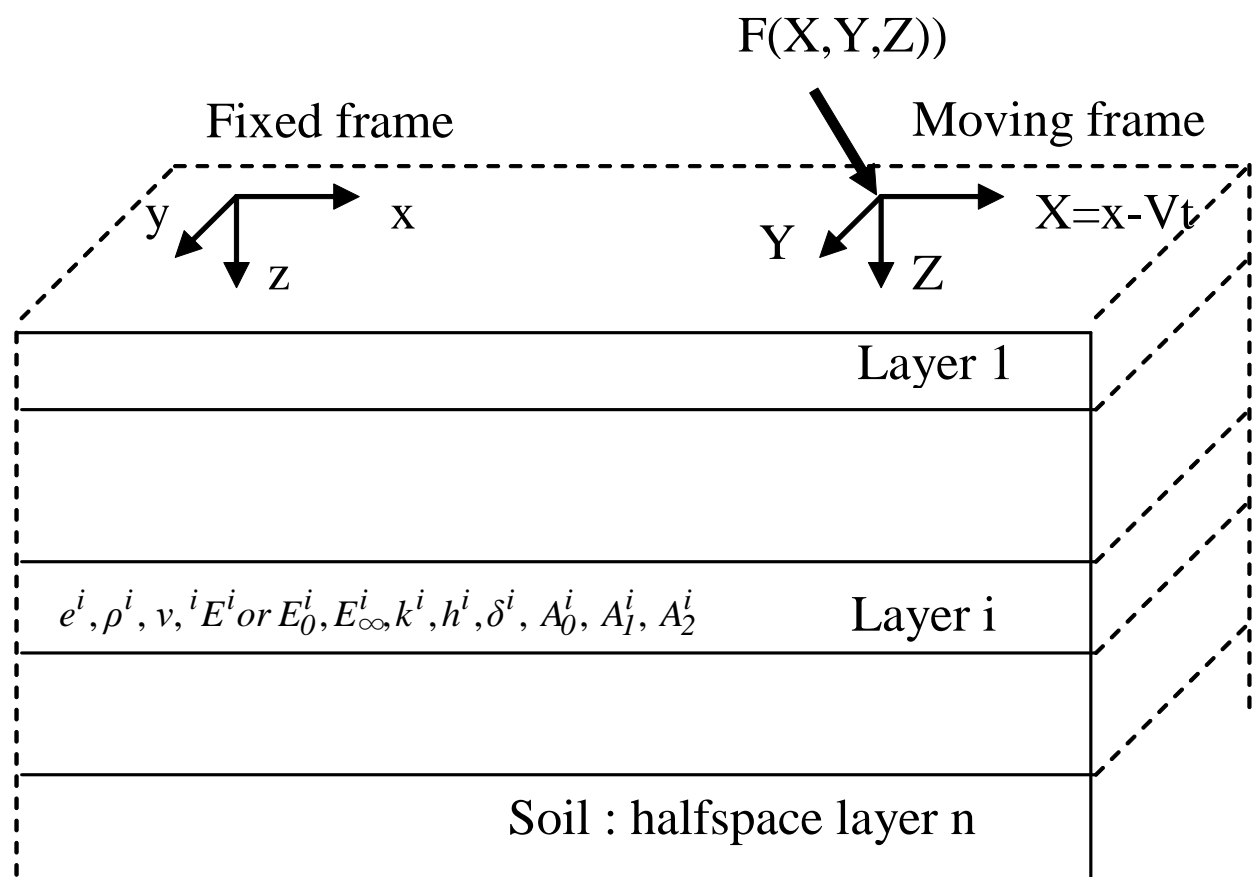

**Figure 1.** *Description of the pavement problem under a moving load*

In the modelling described below, the load pressure can be either punctual or uniformly distributed on a rectangular or an elliptical surface area. In this article layers are assumed to be perfectly bonded. Discussions on the sliding interface condition can be found in (Chupin *et al.*, 2009a; Chupin *et al.*, 2009b). For use in the following developments, let us define *(x,y,z)* as a fixed frame linked up to the medium and *(X,Y,Z)* as a moving frame attached to the load.

**2.1** ***Material behaviour of pavement layers***

In the following, each layer $i$ of the pavement structure is homogeneous (Figure 1). The mechanical behaviour of the soil and the unbound granular materials are assumed to be linear elastic. For these materials, $E^i$ and $\nu^i$ denote the Young modulus and Poisson's ratio of the ith layer, respectively.

On the other hand, the mechanical behaviour of asphalt materials is assumed to be linear thermo-viscoelastic and represented by the five viscoelastic coefficients $E_0^i, E_\infty^i, k_i, h_i, \delta_i$ and the three thermal coefficients $A_0^i, A_1^i, A_2^i$ of the complex modulus of the Huet- Sayegh model (Huet, 1963; Sayegh 1965) for the layer $i$. The Huet-Sayegh model consists in two parallel branches (Figure 2). The first branch is made up of a spring and two parabolic dampers that give the instantaneous and the retarded elasticity of asphalt, respectively. The second one is made up of a spring and it represents the static or the long-term elasticity of asphalt.

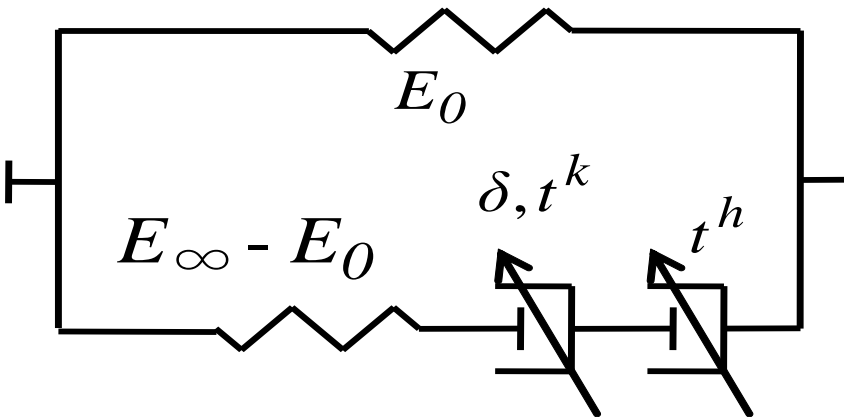

**Figure 2.** *Schematic representation of the Huet-Sayegh rheological model*

By means of parabolic creep laws associated to the two dampers, this rheological model predicts very accurately the complex modulus test obtained for asphalt mixes at different temperatures and frequencies. It has been shown that this simple viscoelastic law is similar to the use of an infinite number of Maxwell branches (Huet, 1999; Heck, 2001; Corté *et al.*, 2004). Note that the coefficients of the Huet-Sayegh model can be determined from experimental tests and by using the free software Viscoanalyse (Chailleux *et al.*, 2006) (see www.lcpc.fr).

Parameter $E_\infty$ is the instantaneous elastic modulus, $E_0$ is the static elastic modulus, $k$ and $h$ are the exponents of the parabolic dampers $(1 > h > k > 0)$, and $\delta$ is a positive adimensional coefficient balancing the contribution of the first damper in the global behaviour. The viscoelastic behaviour is given by the complex modulus [1] that depends on the frequency $\omega$ (with $e^{j\omega t}$ the time variation) and the temperature $\theta$.

$$E\ (\omega,\theta) = E_0 + \frac{E_\infty - E_0}{1+\delta\left(j\omega\tau(\theta)\right)^{-k} + \left(j\omega\tau(\theta)\right)^{-h}} \qquad [1]$$

$\tau(\theta) = \exp\left(A_0 + A_1\,\theta + A_2\,\theta^2\right)$ is a function of temperature and it involves three scalar parameters: $A_0$, $A_1$ and $A_2$.

Poisson's ratio, $\nu^i$, is assumed to be constant and equal to 0.35 for every asphalt material.

## 2.2 *French Asphalt Pavements*

The asphalt pavements mainly used in France could be ranked among four types of pavement structures (SETRA-LCPC, 1997). These structures are listed on Figure 3.

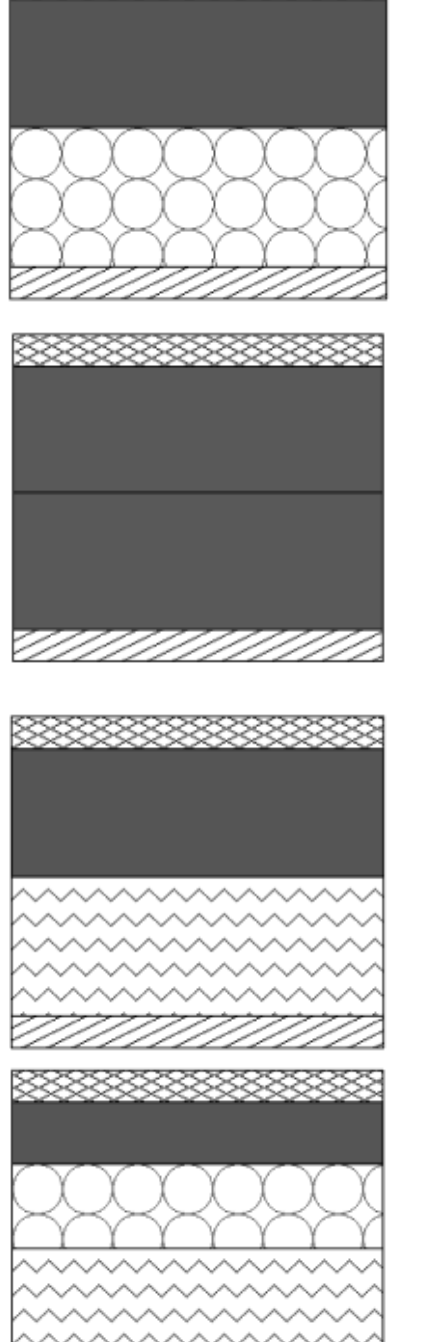

Flexible Pavements
1. Surface course of asphalt materials
2. Base layer of asphalt materials (<15 cm)
3. Unbound granular materials (20 to 50cm)
4. Pavement foundation

Thick asphalt Pavements
1. Surface course of asphalt materials
2. Base layer of asphalt materials (15 to 40 cm)
3. Pavement foundation

Composite Pavements
1. Surface course of asphalt materials
2. Base layer of asphalt materials (10 to 20 cm)
3. Materials treated with hydraulic binders (20 to 40 cm)
4. Pavement foundation

Inverted Pavements
1. Surface course of asphalt materials
2. Base layer of asphalt materials (10 to 20 cm)
3. Unbound granular materials (approx 12cm)
4. Materials treated with hydraulic binders (15 to 50 cm)
5. Pavement foundation

**Figure 3.** *The different French asphalt pavements (SETRA-LCPC, 1997)*

Simulations of flexible pavements have been performed in the past with ViscoRoute 1.0 (Duhamel *et al.*, 2005; Chabot *et al.*, 2006). Moreover, the ViscoRoute software has been used to investigate the effect of horizontal forces of tramway loads on thick asphalt pavement (Hammoum *et al.*, 2009), and the effect of the slip interlayer condition on the mechanical response of a composite pavement (Chupin *et al.*, 2009a). In this article, the interest is focused on thick asphalt pavements and a study is conducted to assess the impact of multiple moving loads on these pavements.

## 2.3 *Types of moving loads*

Different types of moving loads can be considered in pavement design. These relate to single, dual, tandem or tridem tires. To take into account the effects of different configurations of loading, the French design method consists in calculating single or dual loads effects on an elastic pavement. Then, several coefficients are added to predict the effect of tandem and tridem axles (SETRA-LCPC, 1997).

Currently, 40 tons (T) is the French and Europe maximum admissible weight. However, since traffic is increasing, Europe has the desire to increase the total tonnage of freight carried without increasing the maximum weight per axle (11.5T maximum for Europe). Figure 4.a presents one of the commonly used European truck configuration. So, as it is shown in Figure 4.b, to reach 44 and even 50 or 60 tons without inducing further pavement damage, either more axles are required to reduce stress from an axle carrying more weight (Council Directive 96/53/EC, 1996; Council Directive 2002/7/EC, 2002). The French Pavement Design Procedure is under update to examine the prospects of using more tandem axles with the possible use of new wide base tires (455 – 495). Details on these new wide-base tires can be found in (Siddharthan et al., 2002; Wang et al., 2008).

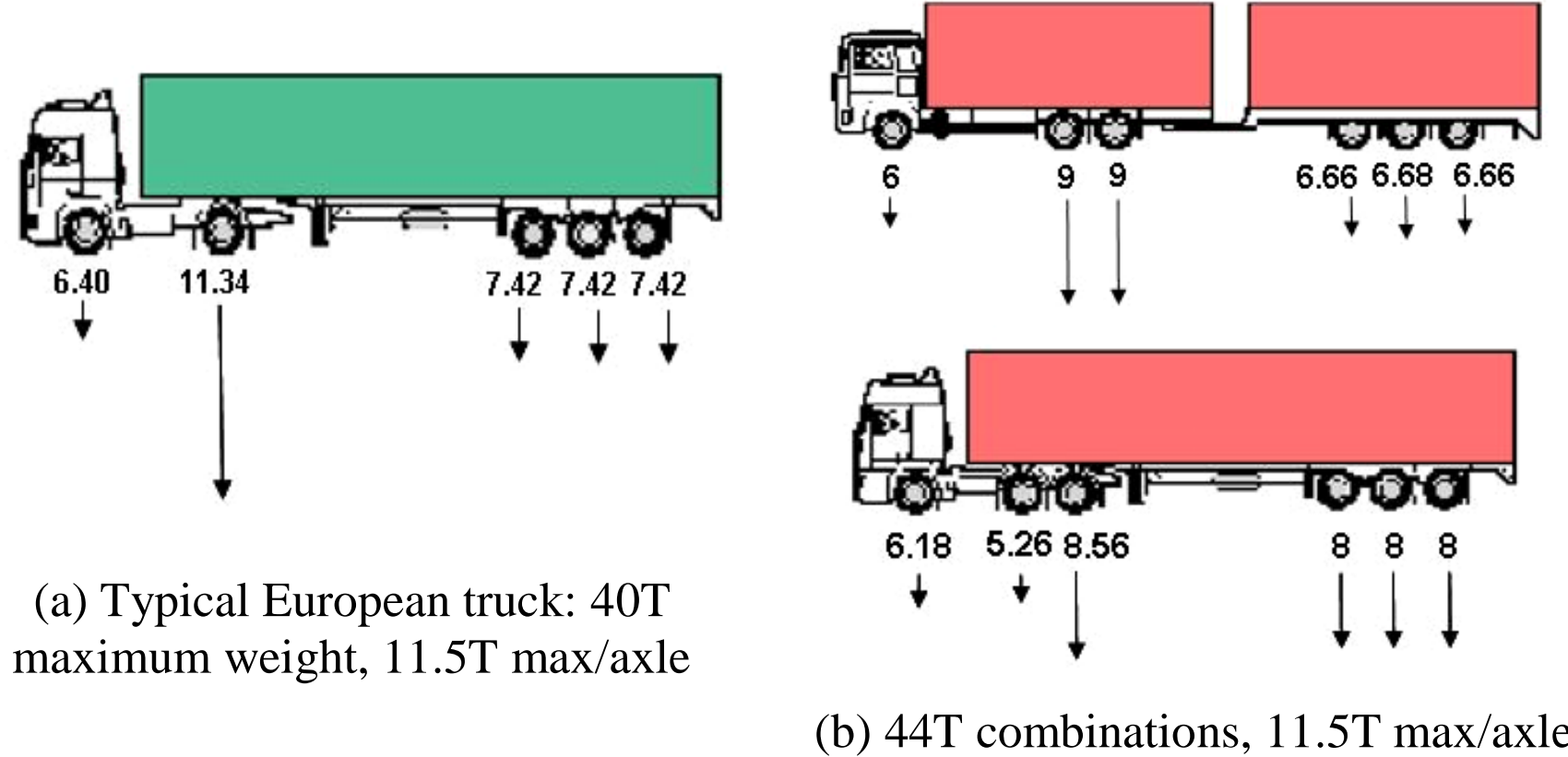

(a) Typical European truck: 40T maximum weight, 11.5T max/axle

(b) 44T combinations, 11.5T max/axle

**Figure 4.** *Typical load configuration of European Truck (http://www.ilpga.ie/public/HGVWeights.pdf)*

In that context and with regard to what have been found concerning aircraft pavement fields (see §6.1), the question to answer is how viscoelastic pavement simulations can help contractors to decide which load configuration is less damaging for pavements.

## 3. Mechanical model

Although inertia forces are generally negligible in pavement applications, to the contrary of Veroad software (Hopman, 1996) and similarly to the 3D-MOVE software (Siddharthan *et al.*, 1998), they can be considered in the modelling which is then governed by the dynamics equations. These equations are solved for each layer in a moving basis attached to the load (Figure 1). To summarize, this section contains only the main equations of the modelling. The interested reader can refer to (Duhamel *et al.*, 2005) for a complete model description.

One shifts from the fixed basis *(x, y, z)*, tied to the medium, to the moving basis *(X, Y, Z)* by making the following change of variable [2]:

$$x = X + Vt; \quad y = Y; \quad z = Z \qquad [2]$$

The elastodynamic equations, with no body forces, expressed in the moving basis $(X,Y,Z)$ reads for each layer [3]:

$$\sigma_{kl,l}(X,Y,Z) = \rho V^2 \frac{\partial^2 u_k(X,Y,Z)}{\partial X^2}, k \in \{1,3\}, l \in \{1,3\} \qquad [3]$$

$(\mathbf{u},\boldsymbol{\sigma})$ denote the displacement and the stress fields, respectively.

### 3.1 ***Solution in the frequency domain and interlayer conditions***

By means Fourier transforms in both $X$ and $Y$ directions, analytical solutions to [3] are computed in the frequency domain in which the viscoelastic constitutive law takes the same form as Hooke's law:

$$\boldsymbol{\sigma}^*(k_1,k_2,Z) = 2\mu_i^*(k_1 V)\boldsymbol{\varepsilon}^*(k_1,k_2,Z) + \lambda_i^*(k_1 V)\, tr\left(\boldsymbol{\varepsilon}^*(k_1,k_2,Z)\right)\mathbf{I} \;, \quad i \in \{1,n\} \qquad [4]$$

The complex Lame coefficients, $\lambda_i^*(k_1 V)$ and $\mu_i^*(k_1 V)$, depend on the complex modulus $E_i^*(k_1 V)$ of the ith layer in the same way as in the elastic case. A Fourier transform applied in the *X* and *Y* directions to [3] combined with [4] yields:

$$\mathbf{A}_i \frac{\partial^2 \mathbf{u}^*(k_1, k_2, Z)}{\partial Z^2} + j\,\mathbf{B}_i \frac{\partial \mathbf{u}^*(k_1, k_2, Z)}{\partial Z} - \mathbf{C}_i \mathbf{u}^*(k_1, k_2, Z) = \mathbf{0} \qquad [5]$$

where *k1* and *k2* are the wave numbers and *j* is the imaginary unit. $\mathbf{u}^*$ is the displacement field expressed in the frequency domain. Matrices $\mathbf{A}_i$ , $\mathbf{B}_i$ and $\mathbf{C}_i$ gather the material properties of layer *i*, $i \in \{1, n\}$. These are given by [6]:

$$\mathbf{A}_i = \begin{pmatrix} c_{si}^2 & 0 & 0 \\ 0 & c_{si}^2 & 0 \\ 0 & 0 & c_{pi}^2 \end{pmatrix} \qquad \mathbf{B}_i = \begin{pmatrix} 0 & 0 & k_1(c_{pi}^2 - c_{si}^2) \\ 0 & 0 & k_2(c_{pi}^2 - c_{si}^2) \\ k_1(c_{pi}^2 - c_{si}^2) & k_2(c_{pi}^2 - c_{si}^2) & 0 \end{pmatrix}$$

$$\mathbf{C}_i = \begin{pmatrix} k_1^2(c_{pi}^2 - V^2) + k_2^2 c_{si}^2 & k_1 k_2 (c_{pi}^2 - c_{si}^2) & 0 \\ k_1 k_2 (c_{pi}^2 - c_{si}^2) & k_1^2(c_{si}^2 - V^2) + k_2^2 c_{pi}^2 & 0 \\ 0 & 0 & k_1^2(c_{si}^2 - V^2) + k_2^2 c_{si}^2 \end{pmatrix} \qquad [6]$$

where $c_{si} = \sqrt{\dfrac{\lambda_i^* + 2\mu_i^*}{\rho_i}}$ and $c_{pi} = \sqrt{\dfrac{\mu_i^*}{\rho_i}}$ denote the dilatation and the shear wave velocities of layer *i*, respectively. Assuming that the displacement $\mathbf{u}^*$ can be written in an exponential form, equation [5] leads to a quadratic equation which is solved by means of eigenvalue techniques (Duhamel *et al.*, 2005). After these mathematical manipulations the solution reads:

$$\begin{aligned}
u_1^*(k_1,k_2,Z) &= k_1\beta_{1i}^- e^{-\kappa_{pi}Z} + \kappa_{si}\beta_{3i}^- e^{-\kappa_{si}Z} + k_1\beta_{1i}^+ e^{\kappa_{pi}Z} - \kappa_{si}\beta_{3i}^+ e^{\kappa_{si}Z} \\
u_2^*(k_1,k_2,Z) &= k_2\beta_{1i}^- e^{-\kappa_{pi}Z} + \kappa_{si}\beta_{2i}^- e^{-\kappa_{si}Z} + k_2\beta_{1i}^+ e^{\kappa_{pi}Z} - \kappa_{si}\beta_{2i}^+ e^{\kappa_{si}Z} \\
u_3^*(k_1,k_2,Z) &= j\kappa_{pi}\beta_{1i}^- e^{-\kappa_{pi}Z} + jk_2\beta_{2i}^- e^{-\kappa_{si}Z} + jk_1\beta_{3i}^- e^{-\kappa_{si}Z} \\
&\quad - j\kappa_{pi}\beta_{1i}^+ e^{\kappa_{pi}Z} + jk_2\beta_{2i}^+ e^{\kappa_{si}Z} + jk_1\beta_{3i}^+ e^{\kappa_{si}Z}
\end{aligned} \qquad [7]$$

In [7], $\kappa_{pi}$ and $\kappa_{si}$ are the longitudinal and the shear wave numbers of layer i. They are defined as follows:

$$\kappa_{pi} = \sqrt{\left(1-\frac{V^2}{c_{pi}^2}\right)k_1^2 + k_2^2} \quad ; \quad \kappa_{si} = \sqrt{\left(1-\frac{V^2}{c_{si}^2}\right)k_1^2 + k_2^2} \qquad [8]$$

The displacement [7] is a function of the horizontal wavenumbers $k_1$ and $k_2$ and of the depth *Z*. Besides, the stress tensor is obtained from the displacement field [7] and the constitutive law [4]. The displacement field depends on the 6 parameters $\left(\beta_{1i}^-, \beta_{1i}^+, \beta_{2i}^-, \beta_{2i}^+, \beta_{3i}^-, \beta_{3i}^+\right)$ that are representative of a layer. Consequently, the solution is completely defined once these parameters have been calculated.

They are determined from the boundary and the interlayer conditions that yield the 6n equations required for the determination of all the parameters. Boundary conditions on the free surface (imposed force vector on the loading area that can be punctual or not) and at infinity (radiation condition) yield 6 equations. The remaining equations are provided by the interlayer relations. In the case of a bonded interface, the continuity relation is used [9]. This relation stipulates that the displacements and the traction vector from both sides of an interface are equal at the Z-coordinate of this interface.

The continuity equation for an interface squeezed between layers i and i+1 reads:

$$\begin{bmatrix} \mathbf{u}^*(k_1,k_2,Z) \\ \boldsymbol{\sigma}^*(k_1,k_2,Z).\mathbf{e}_Z \end{bmatrix}_i = \begin{bmatrix} \mathbf{u}^*(k_1,k_2,Z) \\ \boldsymbol{\sigma}^*(k_1,k_2,Z).\mathbf{e}_Z \end{bmatrix}_{i+1} \qquad [9]$$

Again, solving [9] at all the interfaces within the structure and taking into account boundary conditions enable to compute the unknown coefficients of equation [7]. Once these coefficients have been calculated, the displacement, the strain and the stress fields can be fully determined in the frequency domain (see Duhamel *et al.*, 2005 for more details). The solution in the spatial domain is then obtained by using the Fast Fourier Transform as explained in the upcoming section.

### 3.2 *Solution in the spatial domain*

The Fast Fourier Transform (FFT) is utilized to evaluate the integral that leads to the response in the spatial domain. The FFT is run in two dimensions for all values of $k_1$ and $k_2$ but $k_1$ equal zero. In the latter case, the integrand is singular, though still integrable, and a different method based on Gauss-Legendre polynomials is used (Duhamel *et al.* 2005).

To summarize: the solution obtained in the spatial domain is a component of the displacement, the strain or the stress field at a given z-coordinate in the structure. The solution is thus expressed in a horizontal plan and is computed at many discrete locations in this plan. This solution procedure is implemented in the ViscoRoute kernel that uses the C++ language programming.

### 3.3 *Interpolation of the solution at non-discretized locations*

The solution computed according to the method described above is obtained at discrete locations determined by the number of points used in the FFT. However, one might be interested in getting the solution at non-discretized locations. To accomplish this, the Shannon theorem is employed. Under some assumptions (the considered signal, say *f*, should be composed of frequencies lower than a limit value $\lambda_c$ and its energy must be finite), this theorem leads to an exact interpolation of the solution. In this article, interpolations are performed along lines, i.e. at a given *X* or *Y* discretized location. The Shannon theorem is thus used in only one dimension. In the *X*-direction, it reads:

$$\forall dX \leq \frac{1}{2\lambda_c}, \quad f(X_t) = \sum_{n=-\infty}^{\infty} f(ndX) \frac{\sin \frac{\pi}{dX}(X_t - ndX)}{\frac{\pi}{dX}(X_t - ndX)} \qquad [10]$$

$dX$ is the discretization step in the $X$ -direction and $X_t$ is the location where the interpolation is performed.

## 4. Viscoroute Software

The ViscoRoute software is composed of a computation kernel and a Graphical User Interface (GUI). Two versions of the software Viscoroute have been developed and the difference between each other lies essentially in the GUI. In Viscoroute 1.0 the GUI is programmed in Visual Basic (Duhamel et al., 2005) whereas Python has been used to build up the GUI of Viscoroute 2.0 and then facilitate, during the software installation, all the problems due to different platform support. On the contrary to its first version, ViscoRoute 2.0 offers the possibility to compute the solution with elliptical-shaped loads and several loads directly in the kernel. In this article the second version of ViscoRoute is presented. This version will be downloadable for free on the LCPC website (www.lcpc.fr).

### 4.1 *The Kernel*

As already mentioned the computation kernel is programmed in C++. The kernel of both versions are similar excepted that Viscoroute 2.0 enables to consider multiple moving loads and elliptical-shaped loads. These two versions rest on the modelling described in section 3.

### 4.2 *The Graphic User Interface (GUI) of ViscoRoute 2.0*

The Graphic User Interface (GUI) of ViscoRoute 2.0 was developed by using the Python programming language. To help users to manipulate its French version, a quick overview of its different windows is given below.

The welcome window of the GUI is composed of three spaces: the menu, the toolbar and the workspace. In the menu, it is possible to use the help tool, denoted "aide". The workspace holds three panels that relate to the structure ("Structure"), the loading conditions plus the definition of the computation parameters ("Chargement"), and the visualisation of the results ("Résultats").

A pavement study consists in filling up the GUI for the structure (Figure 5), the loads and the computation requests (Figure 6).

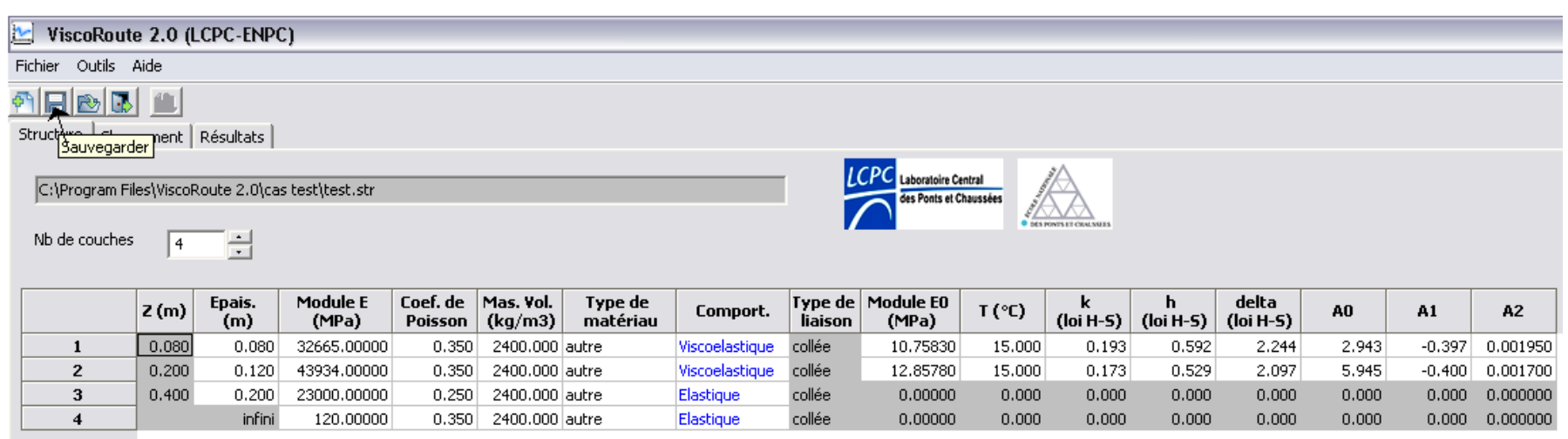

| | Z (m) | Epais. (m) | Module E (MPa) | Coef. de Poisson | Mas. Vol. (kg/m3) | Type de matériau | Comport. | Type de liaison | Module E0 (MPa) | T (°C) | k (loi H-S) | h (loi H-S) | delta (loi H-S) | A0 | A1 | A2 |
|---|---|---|---|---|---|---|---|---|---|---|---|---|---|---|---|---|
| 1 | 0.080 | 0.080 | 32665.00000 | 0.350 | 2400.000 | autre | Viscoelastique | collée | 10.75830 | 15.000 | 0.193 | 0.592 | 2.244 | 2.943 | -0.397 | 0.001950 |
| 2 | 0.200 | 0.120 | 43934.00000 | 0.350 | 2400.000 | autre | Viscoelastique | collée | 12.85780 | 15.000 | 0.173 | 0.529 | 2.097 | 5.945 | -0.400 | 0.001700 |
| 3 | 0.400 | 0.200 | 23000.00000 | 0.250 | 2400.000 | autre | Elastique | collée | 0.00000 | 0.000 | 0.000 | 0.000 | 0.000 | 0.000 | 0.000 | 0.000000 |
| 4 | | infini | 120.00000 | 0.350 | 2400.000 | autre | Elastique | collée | 0.00000 | 0.000 | 0.000 | 0.000 | 0.000 | 0.000 | 0.000 | 0.000000 |

**Figure 5.** *Data for a four layer structure in the ViscoRoute 2.0 GUI*

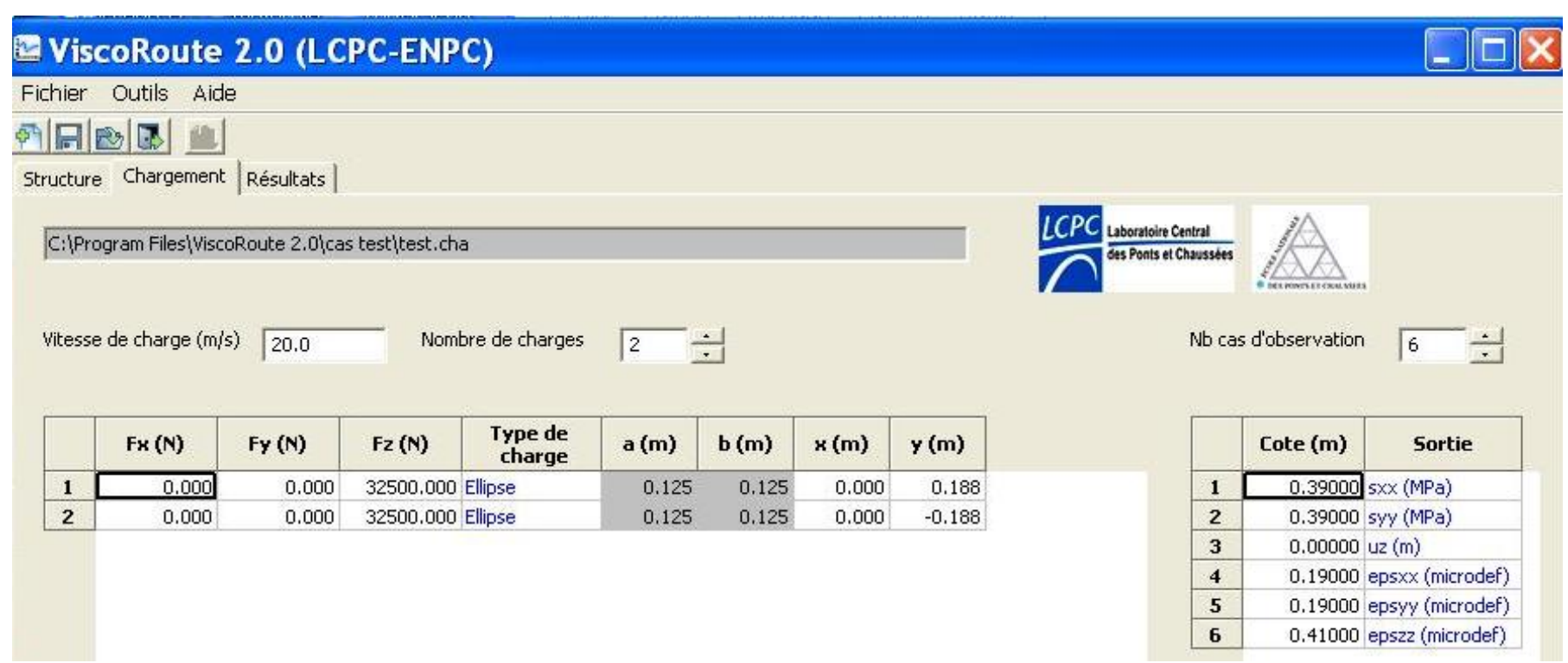

**Figure 6.** *Loading data corresponding to a dual tire and six computation requests in the ViscoRoute 2.0 GUI*

The list of parameters required for one simulation is given in Table1.

**Table1.** *Lexis list of ViscoRoute 2.0 GUI Parameters*

| GUI Parameters | Comments |
|---|---|
| Structure data (Figure 5) | |
| "Nb de couches" | the total number $n$ of layers |
| "z(m)"<br>"Epais.(m)" | the depth of each bottom layer in meter<br>the thickness of layer $i$ $\left(i \in \{1,n\}\right)$ in meter |
| "Module E (MPa)"<br>"Coef. de Poisson"<br>"Mas. Vol."<br>"Type de matériau"<br>"Comport."<br>"Type de liaison"<br>"Module E0 (MPa)"<br>"T (°C)"<br>"k (loi H-S), h (loi H-S), delta (loi H-S)"<br>"A0, A1, A2" | The Young modulus $E^i$ or $E^i_\infty$ in MPa<br>Poisson's ratio coefficient $\left(\nu^i\right)$<br>Density in kg/m$^3$<br>The user can comment the type of material<br>elastic or viscoelastic behaviour of each layer<br>Bonded ("collée") for ViscoRoute 2.0<br>The Huet-Sayegh static elastic modulus $E^i_0$ in MPa<br>Temperature expressed in degree Celsius<br>Parameters $k_i, h_i, \delta_i$ of the Huet-Sayegh model<br>The 3 thermal parameters of the Huet-Sayegh model |
| Load data (Figure 6) | |
| "Vitesse de charge" | The uniform Speed $V$ of the moving load in m/s |
| "Nombre de charge" | Number of applied loads |
| "Fx (N), Fy (N), Fz(N)" | Intensity values of the vector force (N) for each load |
| "Type de charge" | Type of the loaded area: punctual, rectangular or elliptic |
| "a, b" | Half dimensions of the surface load (Figure 7) |
| "x, y" | Coordinates of the load centre |

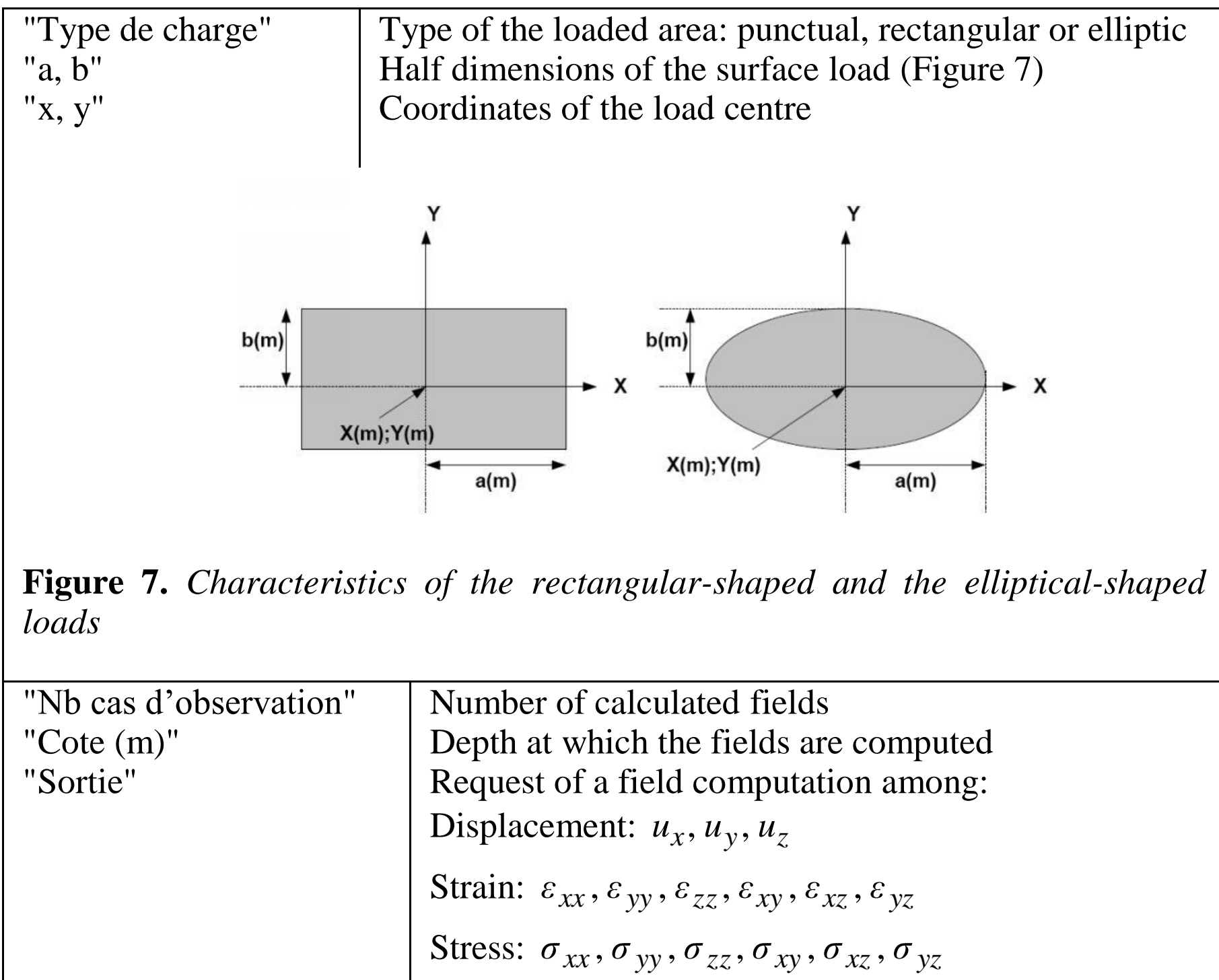

**Figure 7.** *Characteristics of the rectangular-shaped and the elliptical-shaped loads*

| | |
|---|---|
| "Nb cas d'observation" | Number of calculated fields |
| "Cote (m)" | Depth at which the fields are computed |
| "Sortie" | Request of a field computation among:<br>Displacement: $u_x, u_y, u_z$<br>Strain: $\varepsilon_{xx}, \varepsilon_{yy}, \varepsilon_{zz}, \varepsilon_{xy}, \varepsilon_{xz}, \varepsilon_{yz}$<br>Stress: $\sigma_{xx}, \sigma_{yy}, \sigma_{zz}, \sigma_{xy}, \sigma_{xz}, \sigma_{yz}$ |

Once a simulation is done, the computed field can be plotted against *X* (for a given *Y*-coordinate) or *Y* (for a given *X*-coordinate) (Figure 8). Remember that the result of a simulation is a component of a mechanical field calculated at a unique or several imposed *Z*-coordinates. The computed field can also be saved in both text and graphic formats.

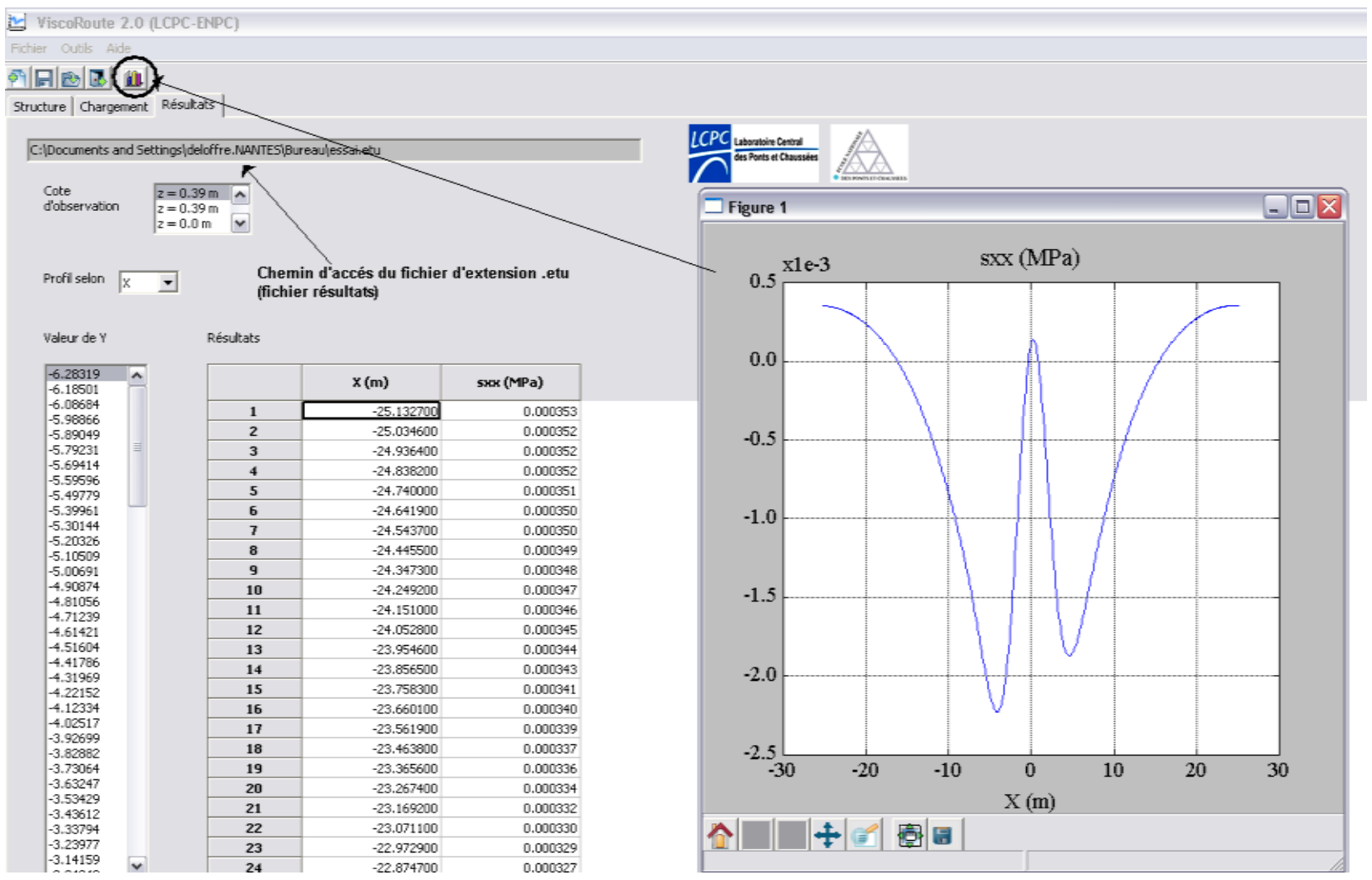

**Figure 8.** *Example of a graphical result in the ViscoRoute 2.0 GUI*

## 5. Validation and comparison with the literature

ViscoRoute has been successfully validated by comparison with analytical solutions derived for an infinite half-space (Chabot *et al.*, 2001) and by comparison to finite element results obtained with the help of the CVCR module of Cesar-LCPC (Heck *et al.*, 1998) in a multilayered case. It has also been used to simulate full-scale experiments (Duhamel *et al.*, 2005).

Moreover, ViscoRoute has been compared with the Veroad® software (Hopman, 1996) since the latter also offers the possibility to take into account the Huet-Sayegh model. The comparison has been conducted for thin and thick flexible pavements which are described in Nilsson et al. (2002) and recalled in Figure 9.

The difference between these two structures only concerns the thickness of the first layer that can be either 0.1m for a thin flexible pavement or 0.2m for a thick flexible pavement. A single moving load (50kN) is applied on the top of the pavements among the *Z*-direction. The circular contact pressure of the load is 800kPa. The velocity of the load is between 10 and 110km/h.

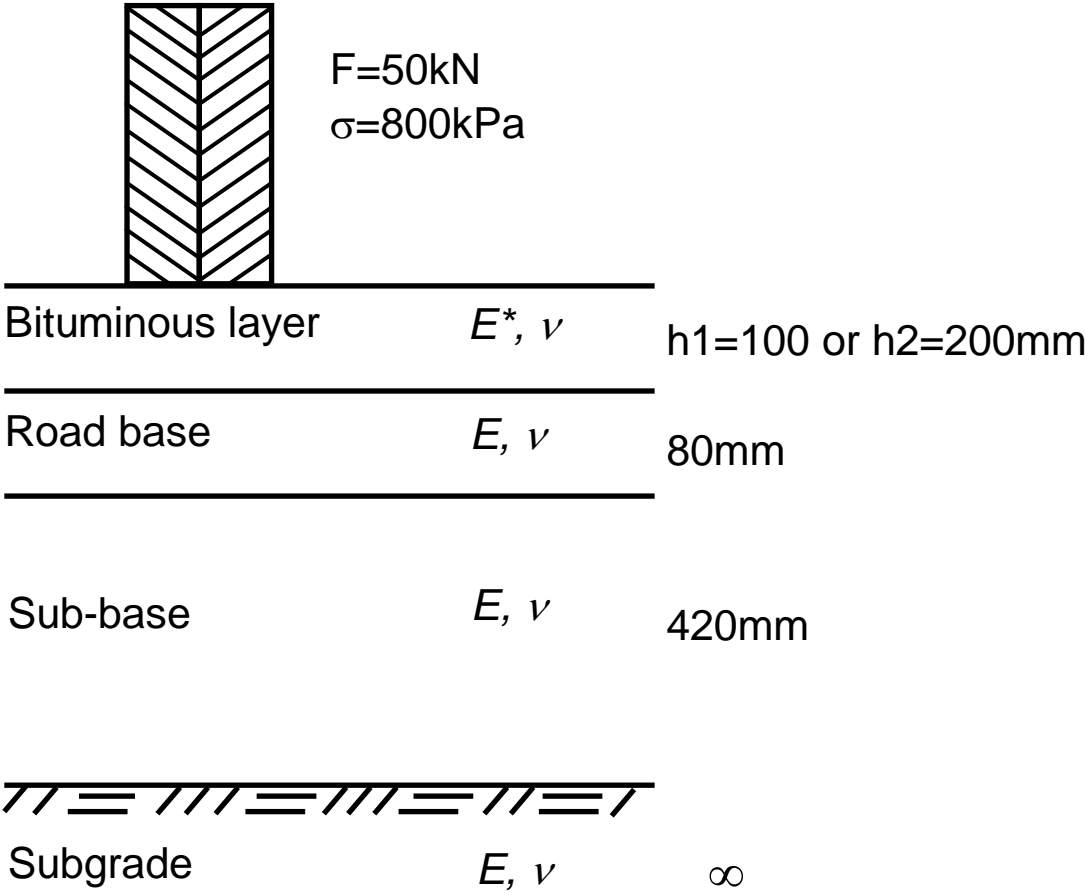

**Figure 9**. *Asphalt pavements studied by Nilsson et al. (2002)*

For each layer, Poisson's ratio is equal to 0.35 and the density is equal to 2100 kg/m$^3$. The first layer is considered as viscoelastic (see Table 2 for the Huet-Sayegh parameters).

**Table 2.** *Average values of the Huet-Sayegh parameters for the asphalt first layer*

| $E_0$ (MPa) | $E_\infty$ (MPa) | δ | k | h | $A_0$ | $A_1$ | $A_2$ |
|---|---|---|---|---|---|---|---|
| 43 | 33000 | 2.550 | 0.269 | 0.750 | -0.86135 | -0.37499 | 0.004534 |

The road base (0.08m in thickness) and the sub-base (0.42m in thickness) are assumed to be elastic and to depend on thermal and moisture characteristics (Table 3) (Nilsson et al., 2002).

**Table 3.** *Young Modulus of elastic materials (Nilsson et al., 2002) (MPa unit)*

| Layer | Winter (<0°C) | Thaw (~0°C) | Summer (20°C) | Autumn (10°C) |
|---|---|---|---|---|
| Road Base | 1000 | 300 | 450 | 450 |
| Sub-base | 450 | 450 | 450 | 450 |
| Subgrade | 1000 | 35 | 100 | 100 |

Similarly to Nilsson *et al.* (2002), the ViscoRoute 2.0 calculations are performed at five different temperatures ranging from -20°C to 20°C. Highest strains are obtained in the transversal direction. So, the comparison between Veroad and ViscoRoute simulations is presented only for the peak values of the transversal strains (Figure 10 and 11).

In Figures 10 and 11, the same tendencies are observed in the computations performed, at different speeds and temperatures, with ViscoRoute 2.0 and Veroad. The inertial forces taken into account in the ViscoRoute modelling seem to not disturb the results for this range of speeds. However, on both asphalt structures, little differences of transversal strain intensity values are found for the highest temperature (20°C) and the slowest speed (10km/h).

These differences are mainly observed for the thin pavement (h1=0.1m) shown in Figure 10. One explanation of these differences could be found in the different ways of computing the solution and introducing the thermal-viscoelastic Huet-Sayegh law. In fact, Veroad introduced the viscoelastic law by means of a linear viscoelastic shear and a linear elastic bulk modulus. ViscoRoute integrates viscoelasticity in a different way by using the complex modulus [1] and assumes that Poisson's ratio is elastic and constant. This last assumption may be inappropriate when viscous effects become important (Chailleux *et al.*, 2009).

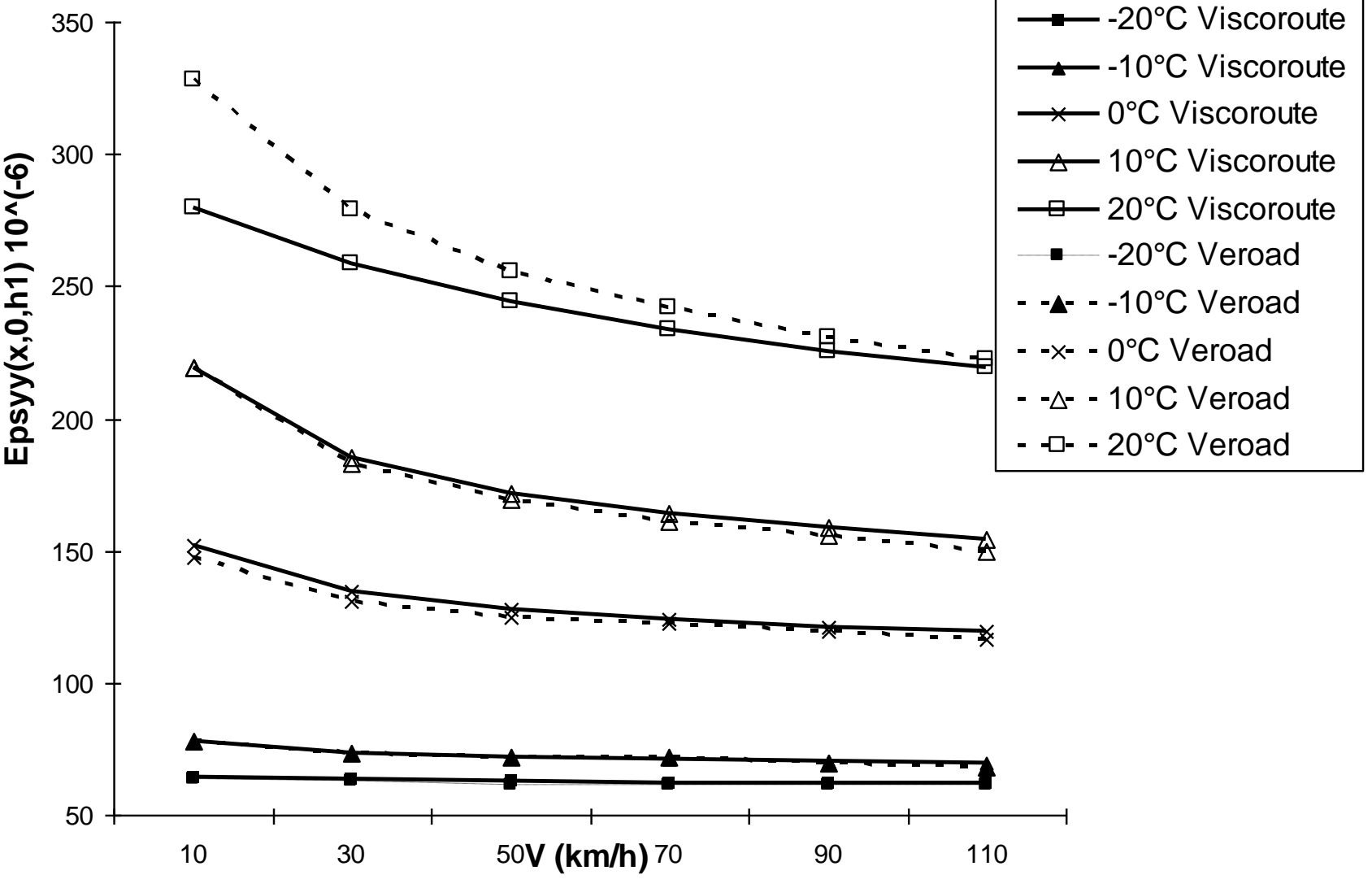

**Figure 10.** *Comparison between Viscoroute 2.0 and Veroad simulations: transversal strain peak values at the bottom of the bituminous layer (h1=0.1mm)*

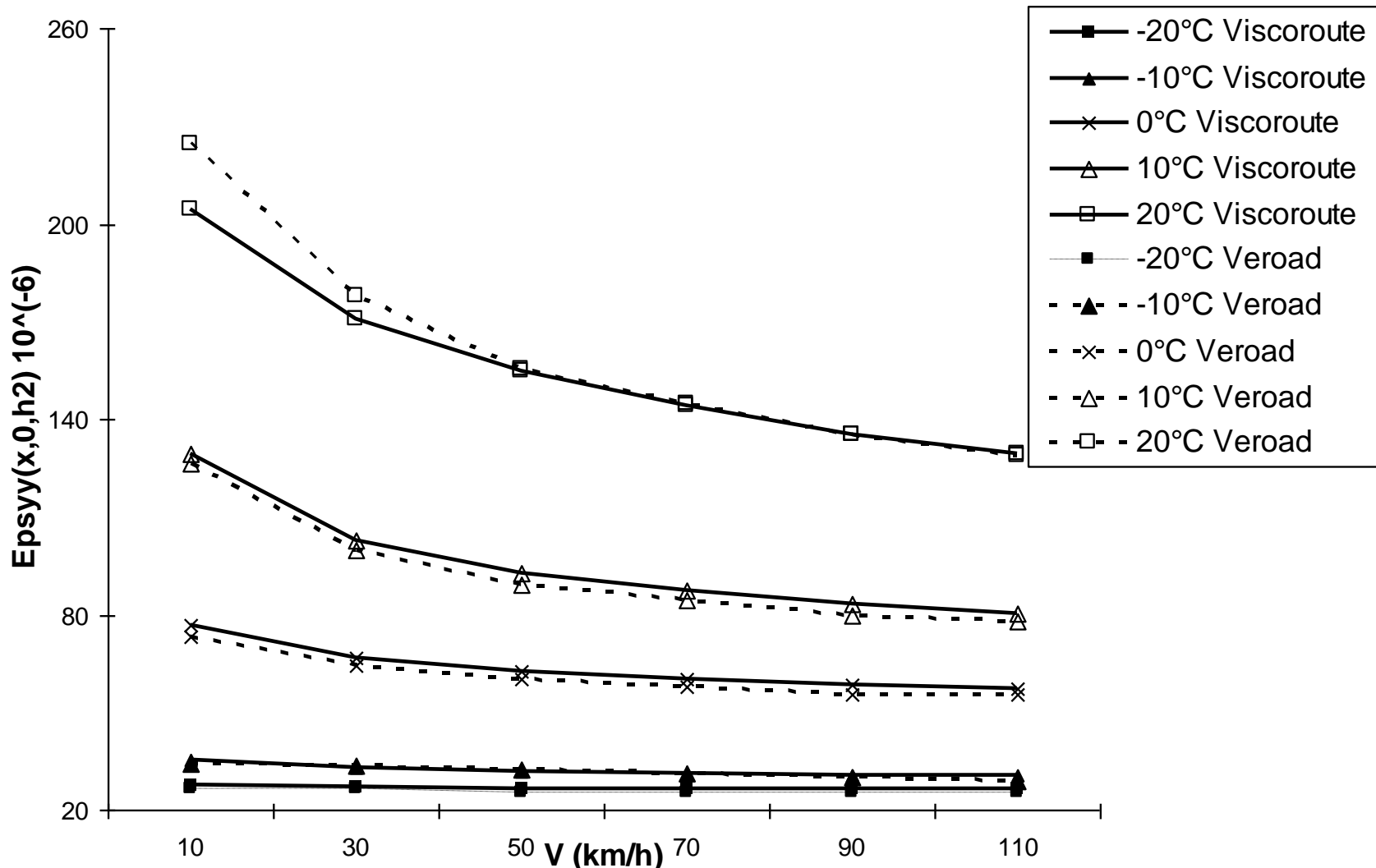

**Figure 11**. *Comparison between Viscoroute 2.0 and Veroad simulations: transversal strain peak values at the bottom of the bituminous layer (h1=0.2mm)*

## 6. Impact of multi-loads on thick asphalt pavement

In this section, the effects of several loads moving on thick asphalt pavement are studied. First, airfield results coming from accelerated pavement test sections are presented. Then, several simulations of dual, tandem and tridem loading configurations are given. Some of the simulations presented herein should help the definition of more realistic signals for fatigue lab tests used in the French Design Method. The aim is to combine these new signals to damage modelling as developed in Bodin *et al.* (2004; 2009) to better predict the fatigue life of asphalt pavement structures subjected to the new generation of trucks.

### 6.1 *Airfield pavement loading results*

In 1999, an A380 Pavement experimental program (PEP) has been done on test sections of thick asphalt pavement (Vila, 2001) (PetitJean *et al.*, 2002). Figure 12 presents typical responses of strain sensors that have been recorded at the bottom of the asphalt layer when submitted to one bogie with four wheels.

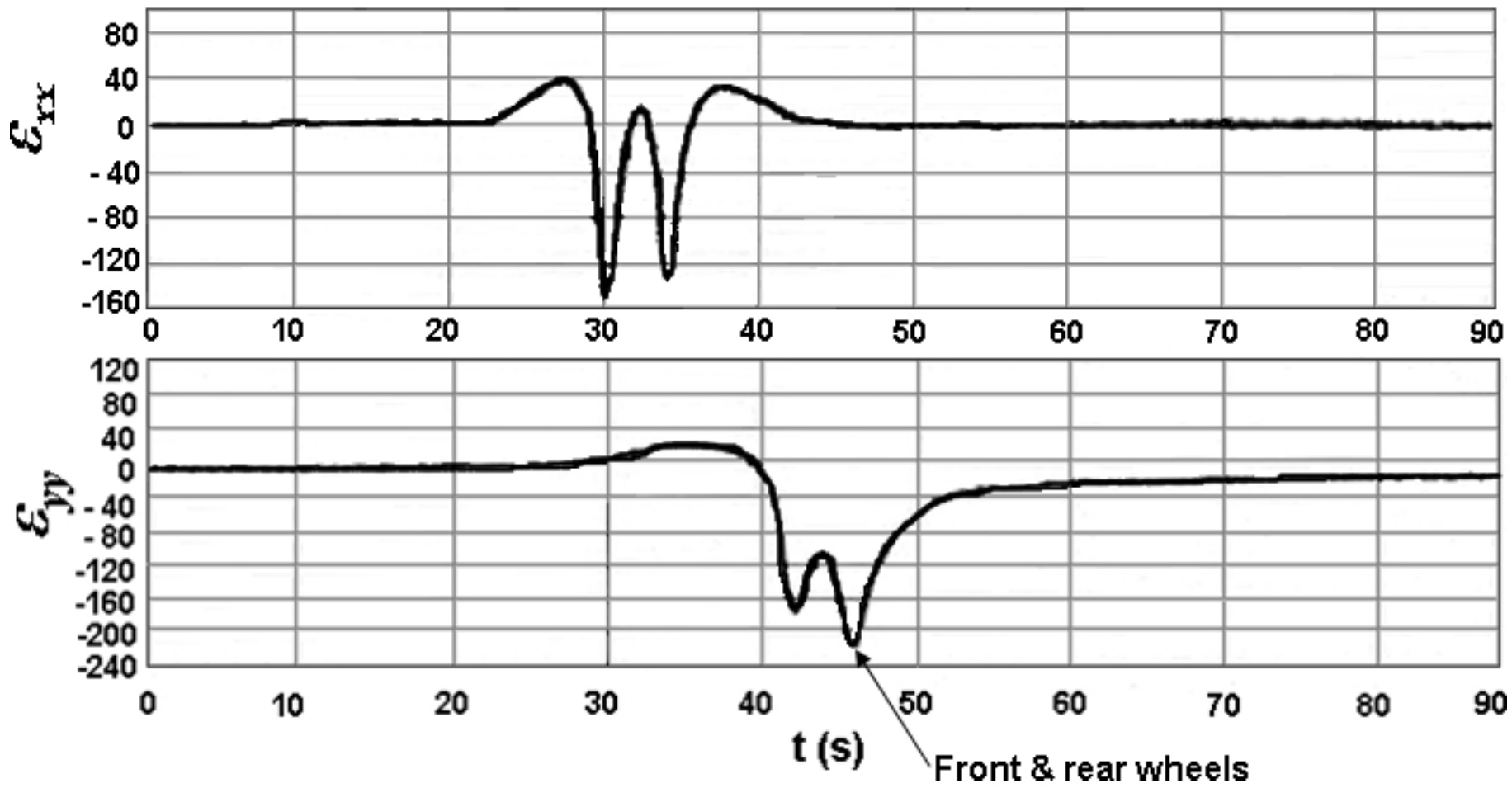

**Figure 12.** *Typical signals of transversal and longitudinal strain gages located at the bottom of the asphalt layer* (PetitJean *et al.*, 2002)

First, it can be observed on Figure 12 that the maximum extension (negative value of strain) is higher in the transversal direction ($\varepsilon_{yy}$) than in the longitudinal one ($\varepsilon_{xx}$). Moreover, the transversal strain signal is strongly asymmetric exhibiting two different peak intensities and it needs some time to return to zero (delay due to viscoelasticity).

To analyze these observations, one of the airfield test section have been studied by Loft (2005). This study is presented hereafter to illustrate the necessity of considering viscoelasticity in the modelling. The pavement test section (Figure 13) is composed of two identical viscoelastic bituminous layers (BB: asphalt concrete and GB: asphalt gravel) whose Huet-Sayegh characteristics are given in Table 4.

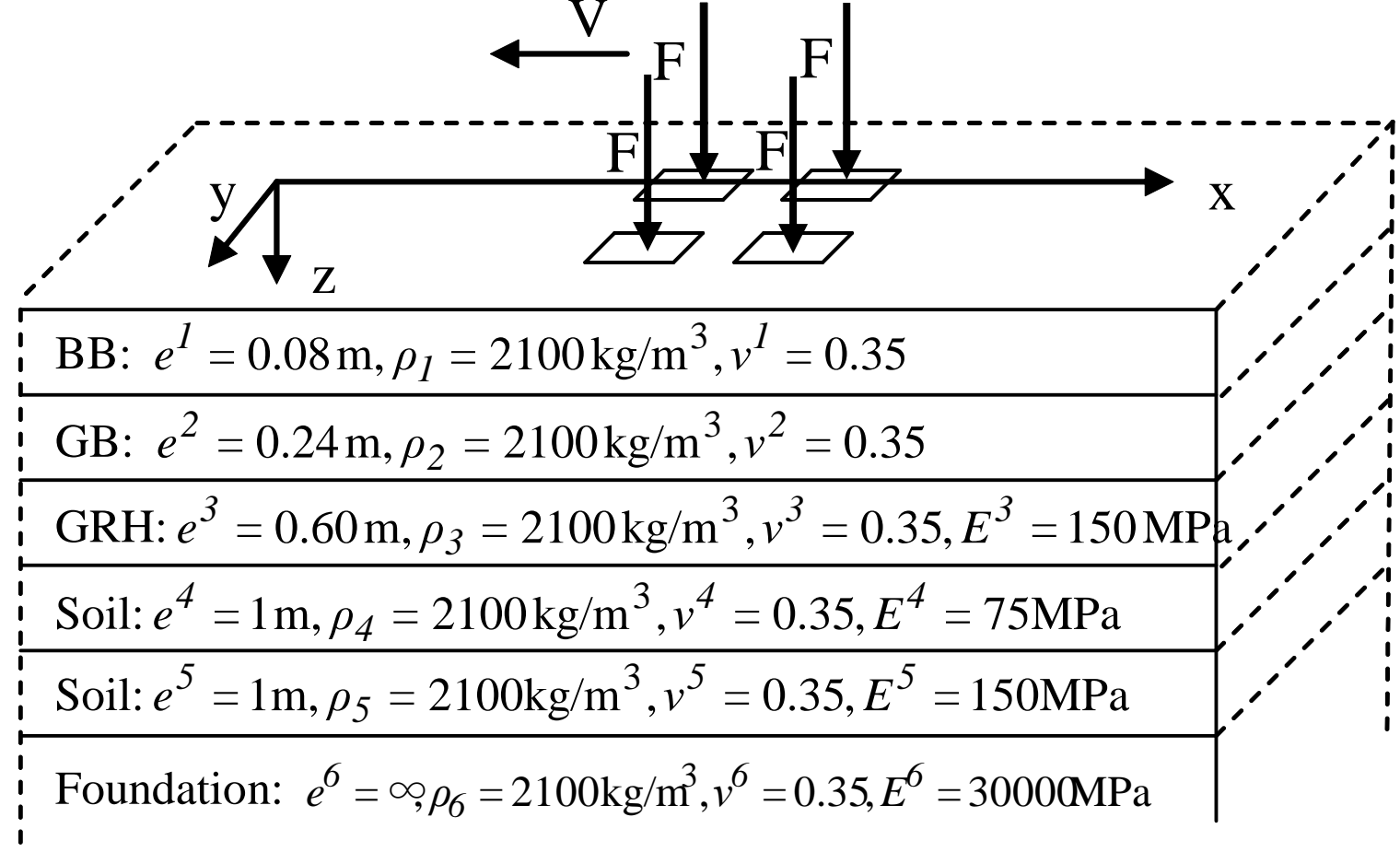

**Figure 13**. *Input data for the analysis of the Aeronautic pavement section*

**Table 4.** *Average values of the Huet-Sayegh parameters of GB and BB layers*

| $E_0$ (MPa) | $E_\infty$ (MPa) | δ | k | h | $A_0$ | $A_1$ | $A_2$ |
|---|---|---|---|---|---|---|---|
| 65 | 30000 | 1.58 | 0.25 | 0.787 | 3.597 | -0.382 | 0.00179 |

The asphalt layers rest on an elastic unbound granular material (GRH: humidify reconstituted crushed gravel) layer and on an elastic soil. The material properties of the elastic layers are obtained by means of backcalculation using finite element simulations (Vila, 2001). The elastic soil is assumed to be composed of two reconstructed subgrade layers resting on a rigid subgrade (Figure 13).

The loads (bogie with four wheels corresponding to the A340 aircraft) applied on the pavement structure move at a constant speed of 0.66m/s. The pressure underneath each individual load (369.6kN) is uniformly distributed on a rectangular-shaped surface (2a=0.56m and 2b=0.40m). The wheelbase of the bogie is of 1.98m along the longitudinal axis ( $x$ ) and 1.40m along the transversal axis ( $y$ ). The thermal sensors positioned within the bituminous layers measured the following thermal distribution: 10.7°C at the top of the pavement section, 10.2°C at a depth of 0.01m, 9.7°C at depths of 0.08 and 0.20m, and 9.3°C at a depth of 0.32m (Vila, 2001). ViscoRoute 1.0 computations have been performed for a single load and the results for the four wheels loading configuration have been obtained by superimposition of the single load case (Loft, 2005). This was possible because of linearity of the constitutive model.

Figure 14 presents the comparison between results obtained by a transversal strain sensor located at the bottom of the Bituminous Gravel (GB), ViscoRoute 1.0 and an equivalent elastic computation ($T_{average}$= 9.7°C, $f$= 0.33 Hz, $E_{eq}$=11670.4MPa) (Loft, 2005). Note that negative values in Figure 14 correspond to extension strains.

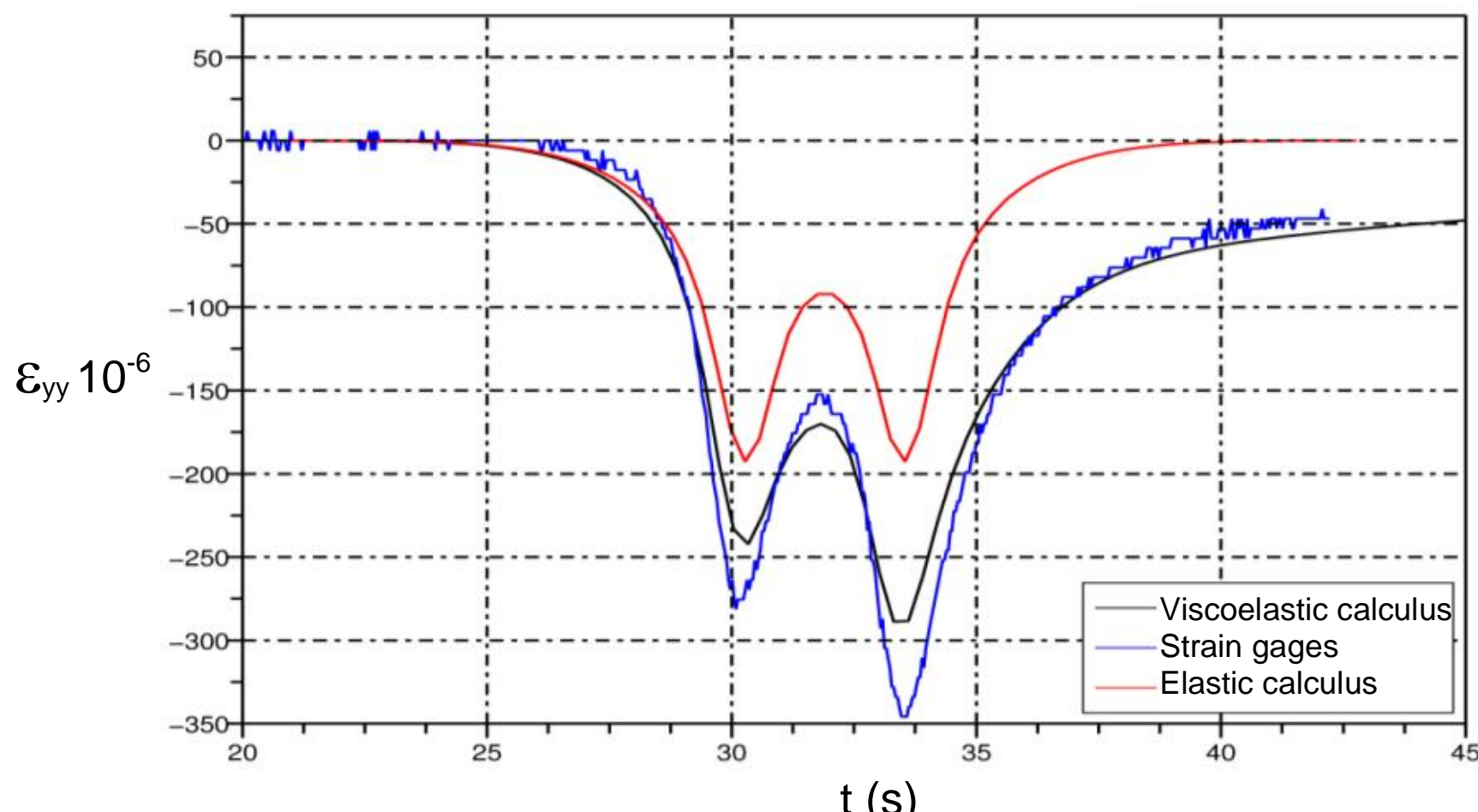

**Figure 14**. *Comparison between elastic computations, ViscoRoute1.0 simulations and transversal strain measurements at the bottom of bituminous layers for a 4-wheels moving load (Loft, 2005)*

In these simulations, the following assumptions on the material properties have been made: similar viscoelastic properties for the BB and the GB layers, and elastic behaviour for other layers. Moreover the location of the strain sensors is assumed to be accurately known.

As shown in Figure 14, the elastic simulation is unsuited to obtain a realistic description of the strains measured at the bottom of the bituminous layer. In particular the peak values are smaller in the elastic simulation than in the measurements. Furthermore, the retardation in the recovery of the transversal strain cannot be predicted by the elastic model. This delay is imputable to viscoelasticity as illustrated by ViscoRoute results that clearly indicate that viscoelasticity of bituminous materials needs to be accounted to get a more realistic simulation of strains produced by aircraft loads moving at low speed on flexible pavements.

### 6.2 *Dual, tandem and tridem effects*

To deepen the previous viscoelastic analysis, the effect of dual, tandem and tridem loads on a thick pavement composed of four layers is studied. The different layers are defined as follows: a surface course of bituminous materials (BB), two base layers of bituminous materials (GB), and a pavement foundation (Figure 15). Table 5 gives the Huet-Sayegh parameters for the three asphalt layers.

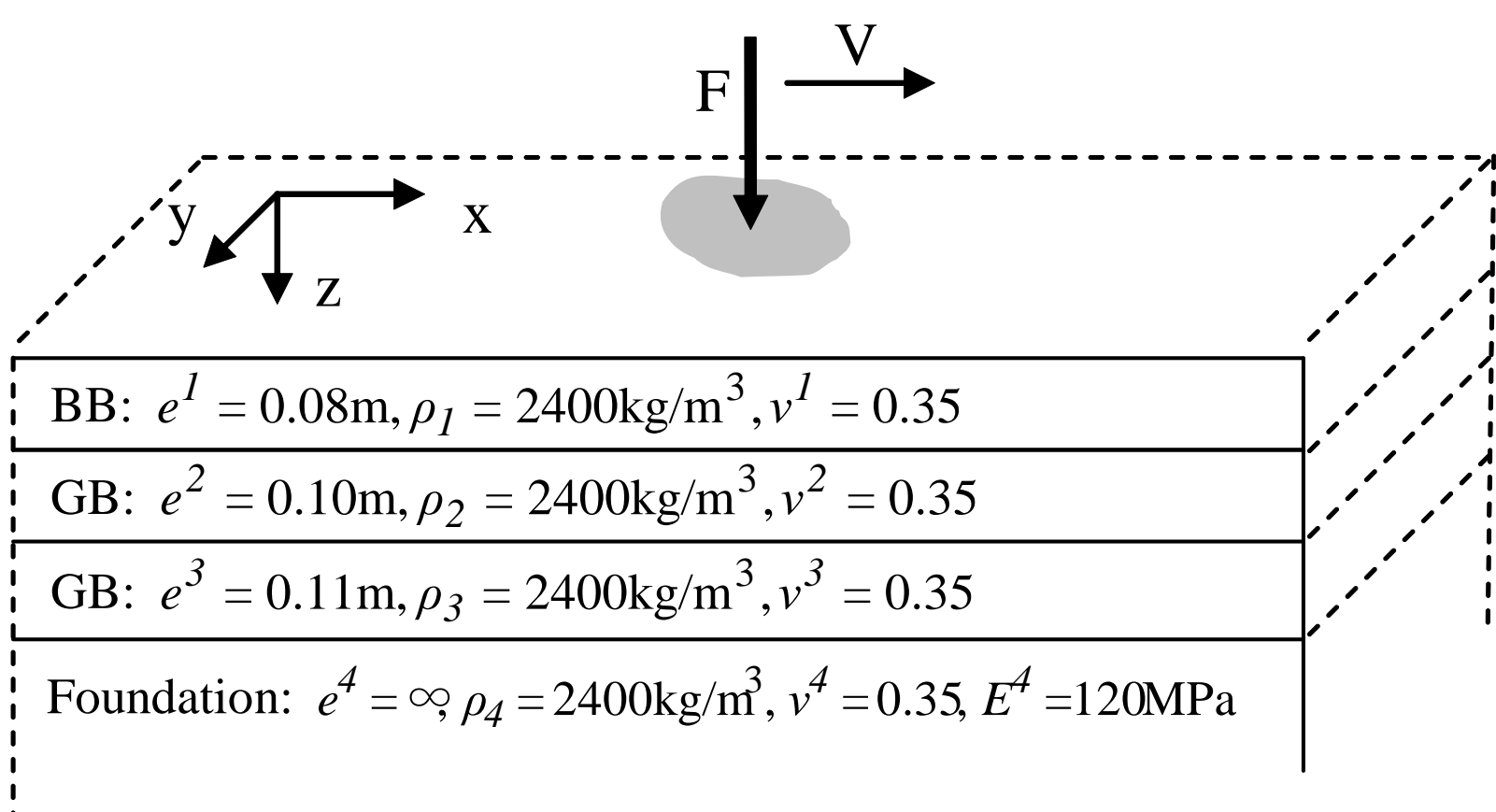

**Figure 15**. *Input data for the analysis of the Thick pavement*

**Table 5.** *Huet-Sayegh parameters for the BB and the GB layers*

| | $E_0$ (MPa) | $E_\infty$ (MPa) | δ | k | h | $A_0$ | $A_1$ | $A_2$ |
|---|---|---|---|---|---|---|---|---|
| BB | 18 | 40995 | 2.356 | 0.186 | 0.515 | 2.2387 | -0.3996 | 0.00152 |
| GB | 31 | 38814 | 1.872 | 0.178 | 0.497 | 2.5320 | -0.3994 | 0.00175 |

Figure 16 presents the characteristics of the contact areas for the different loading configuration.

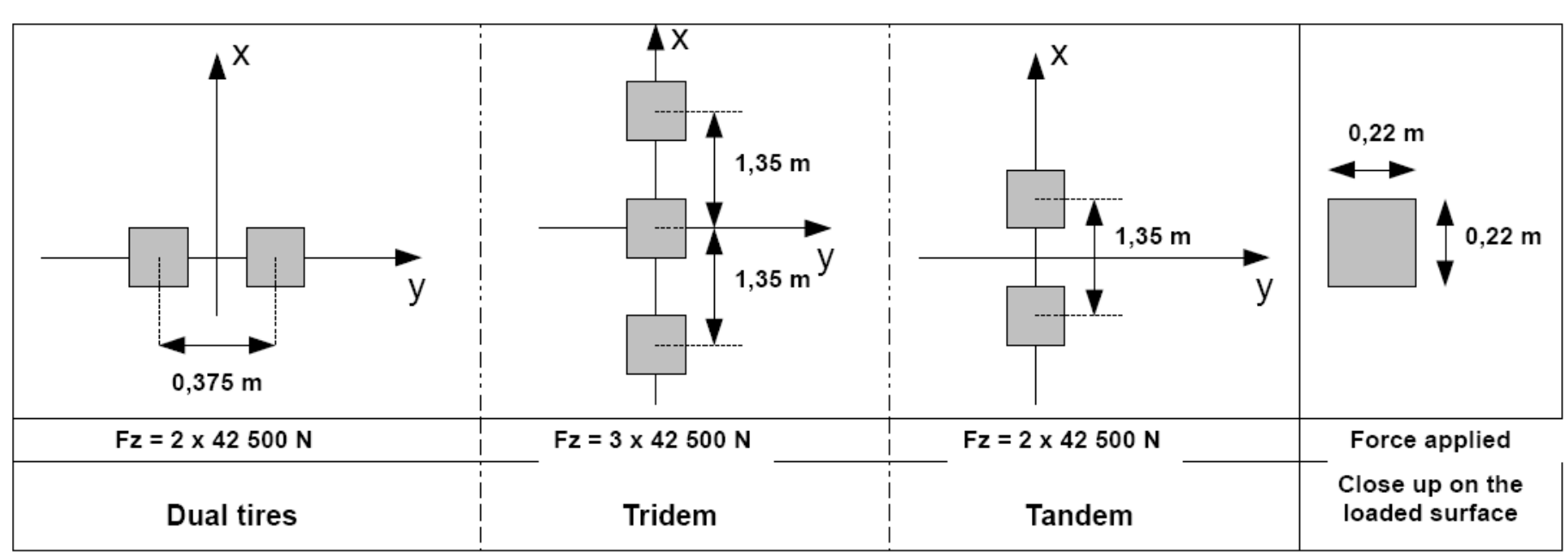

**Figure 16**. *The different type of studied loads*

ViscoRoute computations of the longitudinal and the transversal strains at the bottom of the third layer have been performed for a constant speed of 20m/s and two temperatures (20°C and 30°C). To the contrary of the dual tires (Figure 17), the tandem (Figure 18) and the tridem (Figure 19) configurations lead to higher strains in the transversal direction than in the longitudinal one. A similar trend (with less intensity) would be observed in elasticity.

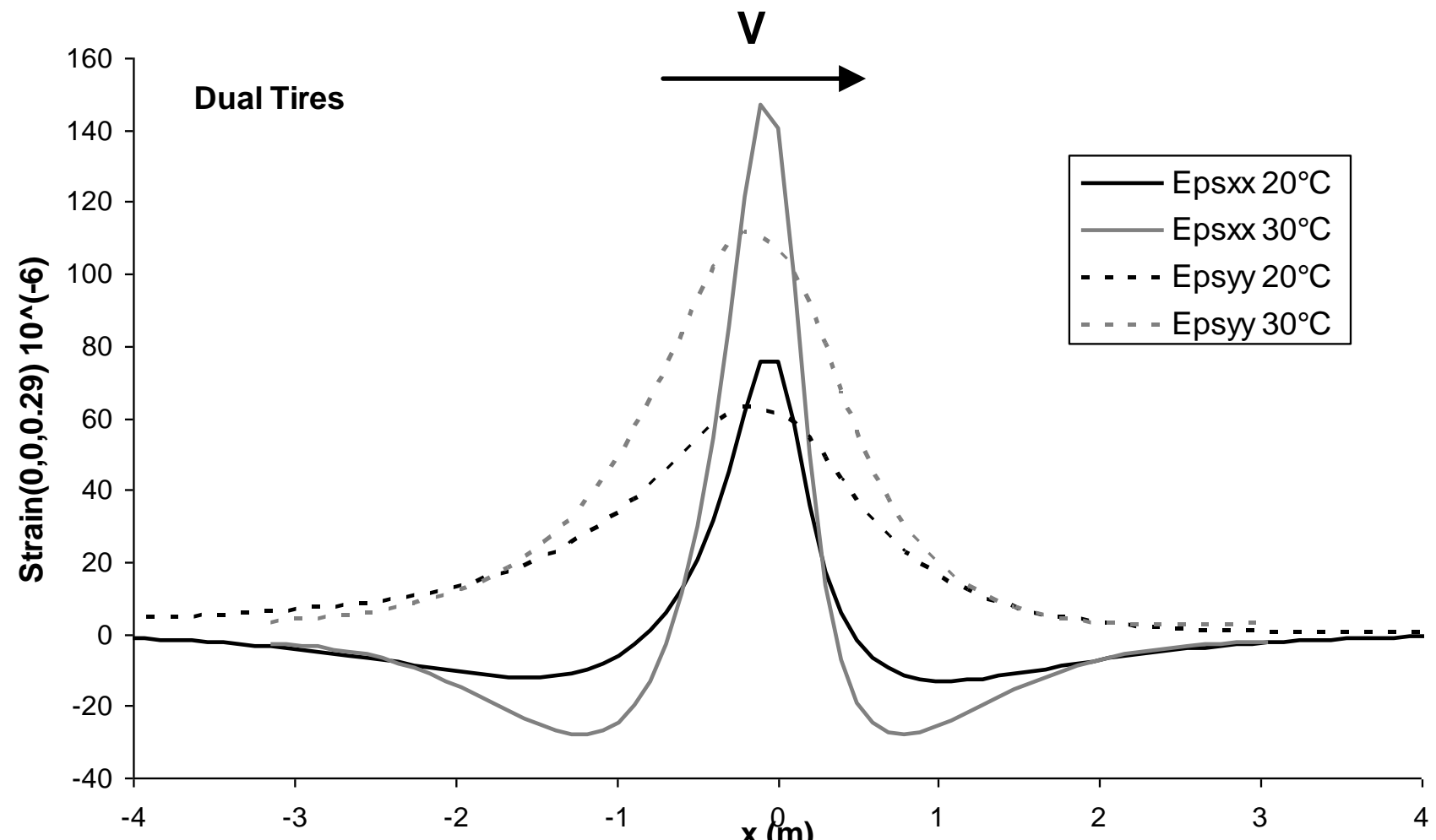

**Figure 17.** *Computed strains at the bottom of 3rd layer for dual tires.*

As already mentioned in section 6.2, the accumulation of transversal strain, which is not predicted at all in elasticity (see Figure 14), is observed in the tandem and the tridem cases (Figure 18 and 19). This effect increases with temperature.

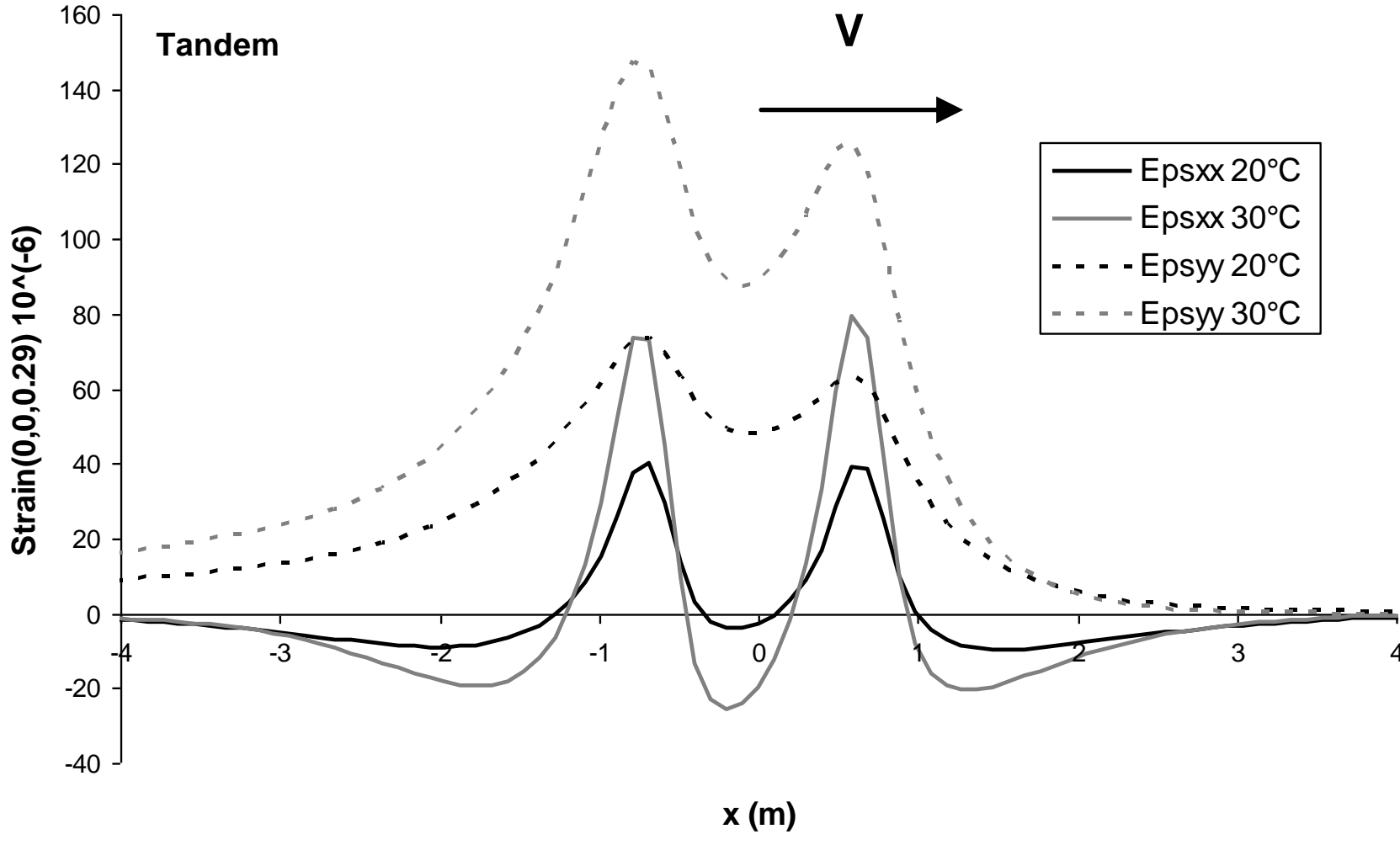

**Figure 18.** *Computed strains at the bottom of 3rd layer for tandem tires*

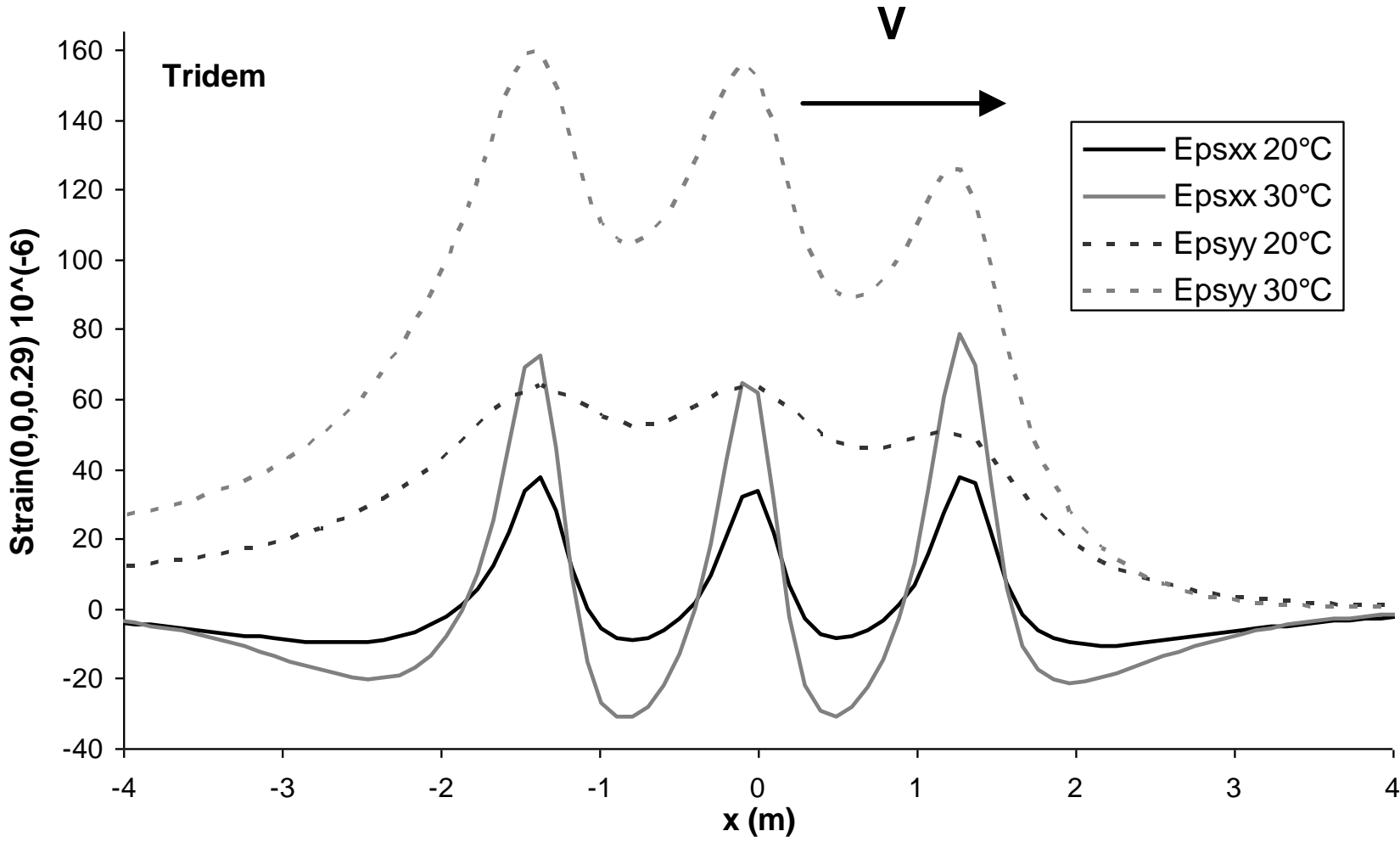

**Figure 19.** *Computed strains at the bottom of 3rd layer for tridem tires*

Finally, the effect of dual tires is compared to tandem tires for equivalent pressure loads (Figure 16). It is observed that the peak value of the transversal strain is quite the same in the dual tire (Figure 17) and the tandem (Figure 18) configurations. Note that in elasticity the magnitude of the deformation for the tandem case would be lower. However, as shown in Figure 20, the computed deflection is higher for the dual tires than for the tandem configuration.

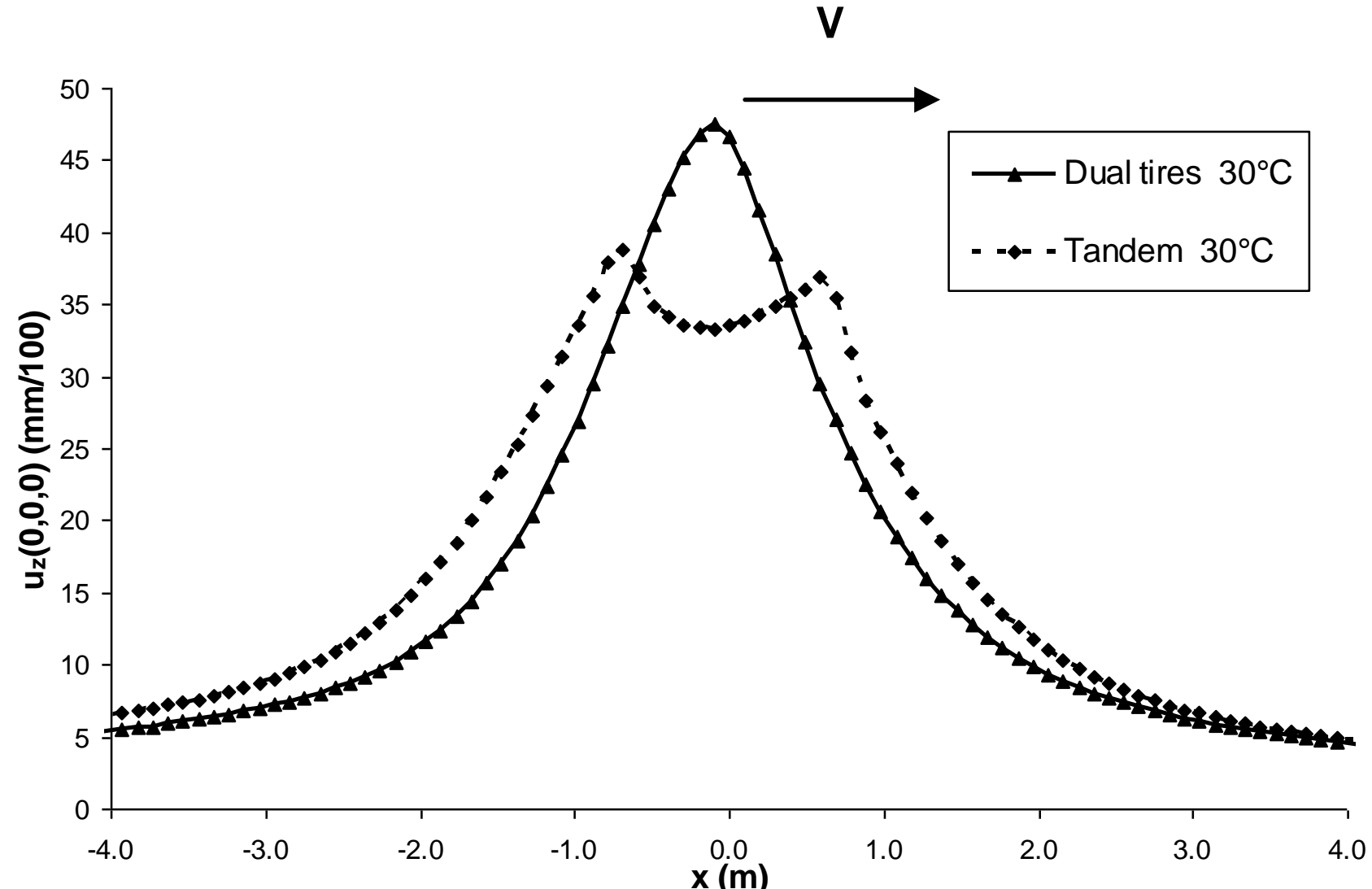

**Figure 20**. *Comparison of the deflection between the dual and the tandem configuration*

As several modelling assumptions have been made in the present study (uniform pressure distribution, fixed wheelbase, linear behaviour for the soil, thick asphalt pavement, moisture effects neglected,…), this last result has to be confirmed. Accelerated pavement projects have already started at LCPC to deepen this study.

## 7. Conclusion/prospects

This article aims at analyzing the influence of moving multi-load effects on the thermo-viscoelastic computed response of asphalt thick pavement structures.

A semi-analytical multi-layered solution using Fast Fourier Transforms and the linear behaviour of the Huet-Sayegh model for asphalt materials has been written in a software called ViscoRoute (Duhamel *et al.*, 2005).

The second version of the software Viscoroute is presented in this paper. ViscoRoute 2.0 enables users to consider multiple moving loads and elliptical-shaped loads. Comparisons with other viscoelastic simulations coming from such similar software as Veroad (Hopman, 1996) have been done and contribute to the validation of the ViscoRoute modelling assumptions.

The accumulation of transversal strains due to multi-loads (such as tandem or tridem configurations) moving on thick asphalt pavements have been successfully simulated with ViscoRoute. This result is in accordance with observations performed during accelerated airfield tests and can not be predicted by an elastic model. If confirmed, this information might be taken into account in the update of the load coefficient used in the French pavement design guide to better predict fatigue life of asphalt pavement with damage modelling.

Finally the latest version of ViscoRoute that enables users to consider perfect slip interlayer relations will be soon available. It is planed to introduce contact laws between layers and non uniform distribution of the load contact pressure. The implementation of complex Poisson's ratio in a way similar to the one described in (Di Benedetto *et al.*, 2007) or (Chailleux *et al.*, 2009) might also be possible.

## 8. Acknowledgements

Authors acknowledge Doctor Viet Tung Nguyen for its contribution to the development of the ViscoRoute 2.0 GUI.

## 9. References

Bodin D., Pijaudier-Cabot G., de La Roche C., Piau J. M., Chabot A., "Continuum Damage Approach to Asphalt Concrete Fatigue Modeling", *Journal of Engineering Mechanics (ASCE)*, Vol. 130, No. 6, 2004, p. 700-708.

Bodin D., Merbouh M., Balay J.-M., Breysse D. Moriceau L., "Experimental study of the waveform shape effect on asphalt mixes fatigue", Proceedings of the *7th Int. RILEM Symp. On Advanced Testing and Characterization of Bituminous Materials*, Rhodes, May 26-28 2009. Vol. 2, p. 725-734.

Chabot A., Piau J.M.,"Calcul semi-analytique d'un massif viscoélastique soumis à une charge roulante rectangulaire", *1ere Conf. Internationale Albert Caquot*, Paris, 2001.

Chabot A., Tamagny P., Duhamel D., Poché D., "Visco-elastic modeling for asphalt pavements – software ViscoRoute", Proceedings of the *10th International Conference on Asphalt Pavements*, Québec, 12-17 August 2006, Vol. 2, p. 5-14.

Chailleux E., Ramond G., Such C., de la Roche C., "A mathematical-based master-curve construction method applied to complex modulus of bituminous materials", *Roads Materials and Pavement Design*, Vol. 7, EATA Special Issue, 2006, p. 75-92.

Chailleux E, de La Roche C, Piau J-M., "Theoretical comparison between complex modulus and indirect tension stiffness of bituminous mixes: influence of the loading law in the indirect tensile test", *Materials and Structures*, 2009, under review.

Chupin O., Chabot A., "Influence of sliding interfaces on the response of a viscoelastic pavement", *6th International Conference on Maintenance and Rehabilitation of*

*Pavements and Technological control, MAIREPAV6*, Torino, 6-10 July 2009a. Edited by Ezio Santagata, Vol 2, p. 675-684.

Chupin O., Chabot A., Piau J.-M., Duhamel D., " Influence of sliding interfaces on the response of a visco-elastic multilayered medium under a moving load", *International Journal of Solids and Structures* , 2010, under review.

Corté J.F., Di Benedetto H., "*Matériaux routiers bitumineux 1: description et propriétés des constituants*", Paris, Lavoisier (traité MIM série Géomatériaux), 2004.

Council Directive 96/53/EC, OJ L 235, September 17 1996, p. 59 amended by Council Directive 2002/7/EC, OJ L 067, 9.3.2002, p. 47

Di Benedetto H., Delaporte B., Sauzéat C., "Three-Dimensional Linear Behavior of Bituminous Materials: Experiments and Modeling", *International Journal of Geomechanics*, Vol. 7, No. 2, March/April 2007, p. 149-157.

Duhamel D., Chabot A., Tamagny P., Harfouche L., "Viscoroute: Visco-elastic modeling for asphalt pavements", *Bulletin des Laboratoires des Ponts et chaussées* (http://www.lcpc.fr/en/sources/blpc/index.php), No. 258-259, 2005, p. 89-103.

Elseifi M.A., Al-Qadi I.L. & Yoo P.J., "Viscoelastic Modeling and Field Validation of Flexible Pavements", *Journal of Engineering Mechanics (ASCE),* Vol. 132, No. 2, 2006, p. 172-178.

Hammoum F., Chabot A., St. Laurent D., Chollet H., Vulturescu B., "Accelerating and Decelerating Effects of Tramway Loads moving on Bituminous Pavement", *Materials and Structures,* 2009, (available on line December 15 2009 , doi: 10.1617/s11527-009-9577-9).

Heck J.V., 2001. "Modélisation des deformations réversibles et permanents des enrobes bitumineux – Application à l'orniérage des chaussées", Ph.D. dissertation, Université de Nantes.

Heck J.V., Piau J.M., Gramsammer J.C., Kerzreho J.P., Odeon H., "Thermo-visco-elastic modelling of pavements behaviour and comparison with experimental data from LCPC test track", *Proc. of the 5th BCRA,* Trondheim, 6-8 July 1998.

Hopman P.C., "VEROAD: A Viscoelastic Multilayer Computer Program", *Transportation Research Record*, Vol. 1539, 1996, p. 72-80.

Huet C., 1963. Etude par une méthode d'impédance du comportement viscoélastique des matériaux hydrocarbonés. Ph.D. dissertation, Université de Paris.

Huet C., "Coupled size and boundary-condition effects in viscoelastic heterogeneous and composite bodies", *Mechanics of Materials*, No. 31, 1999, p. 787-829.

Nilsson R.N., Hopman P.C., Isacsson U., "Influence of different rheological models on predicted pavement responses in flexible pavements", *Road Materials and Pavement Design*, Vol. 3, No. 2, 2002, p. 117-147.

Loft A., "Evaluation de Viscoroute-v1 pour l'étude de quelques chaussées souples", Msc. Dissertation, Dresden University of Technology speciality Urban and Road construction, 2005.

Petitjean J., Fabre C., Balay JM., "A380 flexible pavement experimental program : data acquisition and treatment process, first numerical simulations and material testing", *Proceedings of the FFA Airport Technology Transfer Conference*, Atlantic City, 2002.

Siddharthan R.V., Yao J., Sebaaly P.E., "Pavement strain from moving dynamic 3D load distribution", *Journal of Transportation Engineering,* Vol. 124, No. 6, 1998, p. 557-566.

Siddharthan R.V., Krishnamenon N., El-Mously M., Sebaaly P.E., "Investigation of Tire Contact Stress Distributions on Pavement Response", *Journal of Transportation Engineering*, Vol. 128, No.2, 2002, p. 136-144.

Sayegh G., "Contribution à l'étude des propriétés viscoélastiques des bitumes purs et des bétons bitumineux", Ph.D. dissertation, Faculté des Sciences de Paris, 1965.

Setra-LCPC, "French Design Manual For Pavement Structures", Laboratoire Central des Ponts et Chaussées and Service d'Etudes Techniques des Routes et Autoroutes, 1997.

Sneddon I.N., "The stress produced by a pulse of pressure moving along the surface of a semi-infinite solid", *Rend. Circ. Mat. Palermo*., Vol. 2, 1952, p. 57-62.

Vila B., "Modélisation numérique des structures de chaussées souples en viscoélasticité et première modélisation des chaussées rigides", Msc dissertation, INSA Toulouse, 2001.

Wang H., Al-Qadi I.L., "Ful-depth Flexible Pavement Fatigue Response under Various Tire and Axle Load Configurations", *APT conference*, 2008.